\RequirePackage{fix-cm}
\documentclass{svjour3}                     
\smartqed  
\usepackage[caption=false]{subfig}
\usepackage{floatrow}
\usepackage{graphicx}
\usepackage{natbib,amsfonts,amssymb}
\usepackage[section]{placeins}
\usepackage{times} 

\begin{document}

\title{A study of the 1/2 retrograde resonance }

\subtitle{Periodic orbits and resonant capture}

\titlerunning{}        

\author{ M.H.M. Morais   \and F. Namouni \and G. Voyatzis \and T. Kotoulas }


\institute{M. H. M. Morais, Instituto de Geoci\^encias e Ci\^encias Exatas, Universidade Estadual Paulista (UNESP), Av. 24-A, 1515 13506-900 Rio Claro, SP, Brazil \at
                         \and
F. Namouni \at
Universit\'e C\^ote d'Azur, CNRS, Observatoire de la C\^ote d'Azur, CS 34229, 06304 Nice, France
  \and
G. Voyatzis \at
Aristotle University of Thessaloniki, Department of Physics, 54124, Thessaloniki, Greece
  \and
T. Kotoulas \at
Aristotle University of Thessaloniki, Department of Physics, 54124, Thessaloniki, Greece
}

\date{Received: date / Accepted: date}

\maketitle

\begin{abstract}

We describe the families of periodic orbits in the 2-dimensional 1/2 retrograde resonance at mass ratio $10^{-3}$, analyzing their stability  and bifurcations into 3-dimensional periodic orbits.  We explain the role played by  periodic orbits  in adiabatic resonance capture, in particular how   the proximity between  a stable family  and an unstable  family with a nearly critical segment, associated with Kozai separatrices, determines the transition between distinct resonant modes observed in numerical simulations. Combining the identification of stable, critical and unstable periodic orbits  with analytical modeling, resonance capture simulations and computation of stability maps helps to unveil the complex 3-dimensional structure of resonances. 
\keywords{Resonance. Three-body problem.}
\end{abstract}

\section{Introduction}

The study of retrograde resonances in the restricted 3-body problem was initiated in \citet{MoraisGiuppone2012,MoraisNamouni2013CMDA} with later developments by \citet{MoraisNamouni2016CMDA,NamouniMorais2018DF,NamouniMorais2020}.
There are now examples of retrograde resonances in the Solar System for the period ratios  1/2 and 2/5 with Jupiter (Centaurs (330759) 2008 S0218 and 2006 BZ8), 2/3 with Saturn (Centaur 2009 QY6) \citep{MoraisNamouni2013MNRASL} and 1/1  with Jupiter (asteroid (514107) Ka'epaoka'awela) \citep{MoraisNamouni2017,Wiegertetal2017} and such configurations are likely to exist in extrasolar systems \citep{GayonBois2008,GayonBois2009,LiMustillDavies2019}.
Additionally, high inclination and retrograde orbits currently orbiting our Sun may be of interstellar origin \citep{NamouniMorais2018MNRASL,NamouniMorais2020Centaurs} hence the study of  high inclination and retrograde  resonances is necessary to understand  how these could be captured by the Solar System.

The numerical study on resonance capture at arbitrary inclination by \citet{NamouniMorais2015,NamouniMorais2017,NamouniMorais2018JCAM} showed that retrograde resonances capture more efficiently than prograde resonances and that this effect is more  pronounced for the 1/1 and 1/2 retrograde resonances. 
Moreover, a 3-phase capture mechanism for retrograde resonances was  observed in the numerical simulations by \citet{NamouniMorais2015,MoraisNamouni2016CMDA}. The observed capture stages consist of: 1) libration of the prograde resonant angle at small eccentricity; 2)  Kozai resonance with coupled eccentricity and inclination oscillations   where both prograde and retrograde resonant angles librate; 3) exit of the Kozai resonance and libration of the retrograde resonant angle.

Computation of periodic orbits (POs)  is a powerful tool to unveil the phase space structure in conservative dynamical systems. In the planar circular restricted 3-body problem (CR3BP),  the motion occurs in a 3-dimensional surface embedded in the 4-dimensional phase space. A 2-dimensional section of this surface is called a Poincar\'e map  and POs  are  fixed points on this map which may be stable or unstable.  When there is a dominant central mass (e.g. the case Sun-planet-asteroid),  stable POs correspond to nearly circular orbits or exact resonances between the asteroid and planet's orbital frequencies, while  unstable POs correspond to  separatrices which delimit the regions of quasiperiodic motion around stable POs.  In the  spatial CR3BP, the motion occurs on a 5-dimensional surface embedded in the 6-dimensional phase space thus Poincar\'e maps are not useful  as a visualization tool. However, the families of stable and unstable POs may still be computed  providing information about the resonances in the 3-dimensional (3D) problem.

Until recently, most 3D searches  were performed for prograde motion (inclination less than $90^\circ$) e.g. in \citet{Voyatzisetal2018}  and \citet{AntoniadouVoyatzis2014b}, based on continuation of planar prograde POs. The few searches which were extended to retrograde motion (inclination larger than $90^\circ$)  were also based on continuation of planar prograde POs \citep{KotoulasVoyatzis2005,AntoniadouLibert2019}.
\citet{MoraisNamouni2019}  computed for the 1st time the  families of stable POs for  the planar retrograde 1/1 resonance and obtained the spatial families which bifurcate from them.
 Subsequently, \citet{KotoulasVoyatzis2020a} computed the families of planar POs for interior retrograde resonances with Jupiter  (i.e.\ with semimajor axis smaller than the planet's)   and more recently the same authors  computed the families of  3-dimensional symmetric POs for the  exterior 1/2,  2/3 and 3/4 resonances with Neptune  (i.e.\ with semimajor axis larger than the planet's) based on continuation from planar prograde and retrograde POs \citep{KotoulasVoyatzis2020b}. \citet{MoraisNamouni2019}  explained how the families of POs for the retrograde 1/1 resonance determine resonant capture, in particular the 3-phase mechanism observed in the simulations \citep{NamouniMorais2015,MoraisNamouni2016CMDA}. The purpose of the current article is to extend the study of \citet{MoraisNamouni2019} to the exterior 1/2 retrograde resonance, focusing on explaining the 3-phase capture mechanism and on unveiling the resonance structure associated with the PO families. This has not been addressed in previous studies.

We performed  a search of periodic orbits  for the retrograde 1/2 resonance in the CR3BP at mass ratio $\mu=10^{-3}$ following the methodology described in Sect.~2.  In Sect.~3  we  describe the families that exist  in the 2D problem and  their bifurcations to the 3D problem. In Sect.~4  we show  examples of adiabatic capture in resonance, the connection with the periodic orbit (PO) families computed previously and predictions from analytical models.   The conclusions of this study  are presented in Sect.~5.

\section{Computation of periodic orbits}

Periodic orbits (POs) of period $T$ satisfy the periodicity conditions $\bar{X}(\bar{X}_0,T)=\bar{X}_0$ where ${X}=(x,z,\dot{x},\dot{y},\dot{z})$ is the phase space vector on the surface of section $y=0$. 
POs  are associated with resonances and may be classified as symmetric, when the critical angle is 0 or $180^\circ$, or asymmetric,  when the critical angle takes a different value . In the former case they are classified as \citep{ZagourasMarkellos1977}: type (a) if symmetric with respect to the $(x,z)$ plane ($\dot{x}(\bar{X}_0,T)=\dot{x}_0=0$, $\dot{z}(\bar{X}_0,T)=\dot{z}_0=0$); or  type (B)  if symmetric with respect to the $x$ axis 
($\dot{x}(\bar{X}_0,T)=\dot{x}_0=0$, $z(\bar{X}_0,T)=z_0=0$). Type (a) and type (B) SPOs  are also known as  F-type and  G-type \citep{KotoulasVoyatzis2005}.

It is known that the planar retrograde 1/2 resonance is a 3rd order resonance  \citep{MoraisGiuppone2012,MoraisNamouni2013CMDA}.
Accordingly,  planar periodic orbits associated with the retrograde 1/2 resonance  have multiplicity 3, i.e. they intersect the Poincar\'e map $y=0$ with the same sign for $\dot{y}$, 3 times per period \citep{MoraisNamouni2013CMDA}. Therefore we restrict our search to POs of multiplicity 3. We compute the circular family of POs with the same multiplicity since from this there may be bifurcations to resonant planar POs  \citep{Hadji1988,Kotoulas&Hadji2002} and  3D POs  \citep{Henon1973,AntoniadouLibert2019,
MoraisNamouni2019,KotoulasVoyatzis2020b}. 

The numerical integration of the CR3BP equations of motion and associated variational equations  were done using the Bulirsch-Stoer algorithm. Distance and time were scaled by the planet's semimajor axis and orbital period. The computations for an individual test particle were stopped when the distance to a massive body was within its physical radius (taken equal to the Sun's and Jupiter's radius). They were also stopped when the heliocentric distance exceeded 10 times the planet's semi-major axis.  

Standard algorithms for computation of POs were used \citep{ZagourasMarkellos1977,Howell1984,Voyatzisetal2018}. These consist of a predictor-corrector scheme where: 1)  a guess  PO, $\bar{X}_0$, is followed until the 3rd  intersection with the surface of section ($|y|<\epsilon_0$) occurs; 2) the periodicity condition  $||\bar{X}-\bar{X}_0||<\epsilon$ is checked; 3) if satisfied  the PO was found and a nearby guess  PO is followed; otherwise a differential correction is applied to the guess PO and the scheme is repeated. Once the PO is found the eigenvalues of the  monodromy matrix $\Delta(T)$ are computed in order to decide if it is stable or unstable  \citep{Hadji2006book}.  As the equations of motion  are symplectic, $|\Delta(T)|=1$, hence these eigenvalues  appear as reciprocal pairs. They may form real or complex conjugate pairs. One pair of (trivial) eigenvalues is unitary  \citep{Hadji2006book}.  Stability occurs if the other pair(s) are complex conjugate on the unit circle.

For planar POs,  2D and vertical stability indexes  may be computed \citep{Hadji2006book,Henon1973}. Stable planar periodic orbits have 2D stability index $-2<k_{2}<2$. Change of stability occurs when $|k_2|=2$ (i.e.\ when the non-trivial pair of eigenvalues is  real on the unit circle)  which is often associated with bifurcation of a new family of POs  \citep{Hadji2006book}.
Motion  around stable 2D periodic orbits is maintained when there are small deviations out of the plane only if the vertical stability index $-2<k_{3}<2$. When $k_3=2$ (vertical critical orbit or {\it vco}) a bifurcation into a new family of  3D periodic orbits with the same multiplicity may occur \citep{Henon1973,Ichtiaroglou1980}.

Generally, to find planar  and 3D POs, it was  sufficient to use double precision arithmetics with parameters 
$\epsilon_0=10^{-13}$ and $\epsilon=10^{-12}$.
To monitor  the POs computations we checked that  $|\Delta(T)|=1$  with at least 10 significant digits. Stability was checked by  explicitly computing the eigenvalues of $\Delta(T)$ and unstable nearly critical motion (near the transition to stability) was confirmed by computing the  chaos indicator MEGNO \citep{MEGNO}. In order to follow the unstable families to termination and to decide on the stability of nearly critical families it was sometimes necessary to use quadruple precision arithmetics.

\section{The 2D families and bifurcations into 3D }

\citet{MoraisNamouni2013CMDA}  showed that  the relevant resonant argument for the planar retrograde 1/2 resonance in the CR3BP is $\phi^*=2\lambda^*+\lambda^*_p-3\varpi^*$. The modified (retrograde) $*$ angles are related to the standard (prograde) angles by the relations $\lambda^*=\lambda-2\Omega$, $\lambda^*_p=-\lambda_p$, $\varpi^*=\varpi-2\Omega=\omega-\Omega$               where  $\lambda$, $\Omega$, $\varpi=\omega+\Omega$ and $\omega$ are the test particle's mean longitude, longitude of ascending node, longitude of pericenter and argument of pericenter, and $\lambda_p$ is the mean longitude of the planet.\footnote{\citet{MoraisNamouni2013CMDA}  showed that the retrograde resonant angles may be obtained from the standard prograde disturbing function by applying a canonical transformation  $\lambda_p^*=-\lambda_p$, $\omega^*=\omega-\pi$, $\Omega^*=-\Omega-\pi$ which is equivalent to inverting the planet's motion hence swapping  ascending and descending nodes.}
 Therefore, we may also write 
$\phi^*=2\lambda-\lambda_p+2\Omega-3\varpi=2\lambda-\lambda_p-\Omega-3\omega$. Note that D'Alembert rule is obeyed when using either modified or standard angles.

\subsection{Planar POs}

There are 2  planar resonant modes, corresponding to the  libration centers: $\phi^*=0$  when the pericenter is interior and at conjunction (aligned) with the planet's orbit; $\phi^*=180^\circ$ when the pericentre is  interior or exterior and at opposition (anti-aligned) with the planet's orbit \citep{MoraisNamouni2013CMDA,MoraisNamouni2016JCAM}.

\begin{figure}
  \centering
   \includegraphics[width=\textwidth]{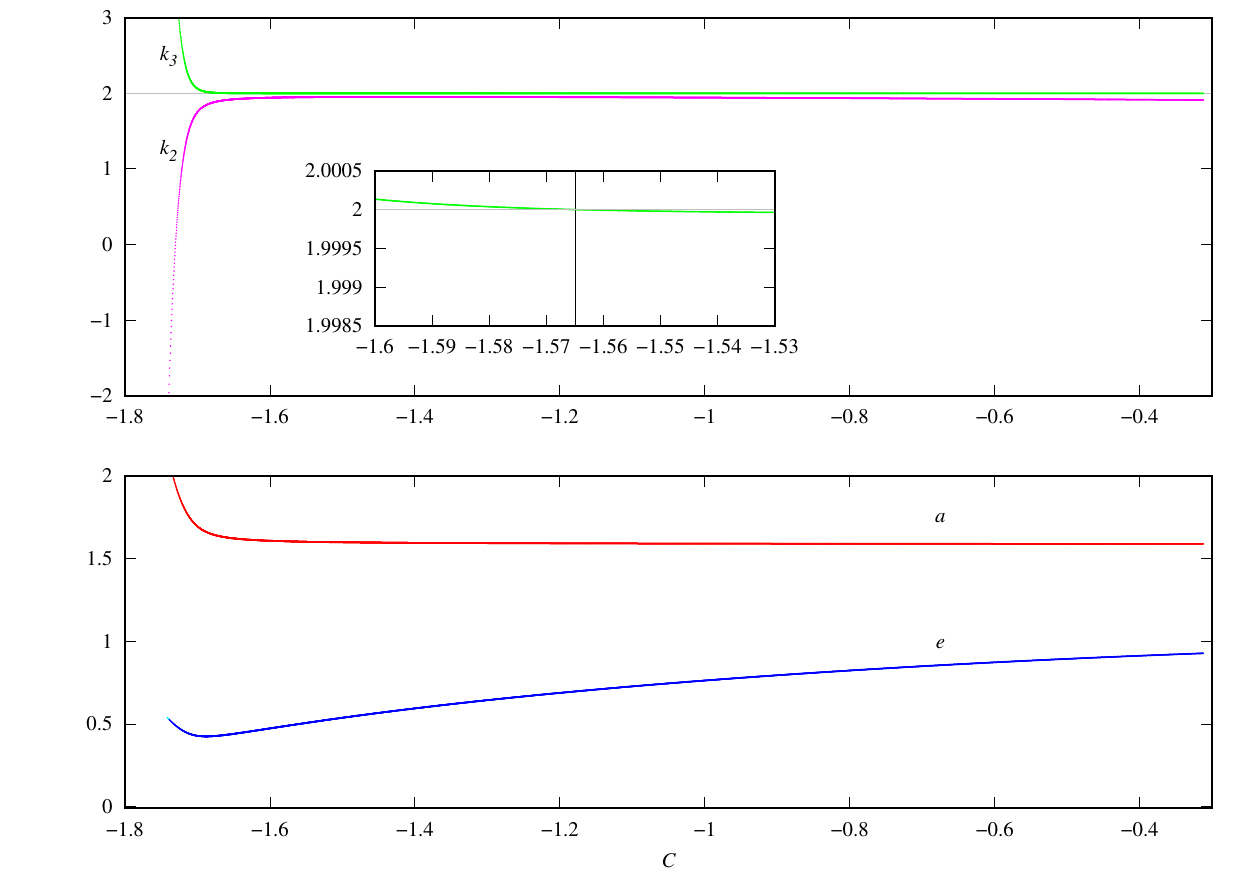}
 \caption{The SPO family corresponding  to the resonance center $\phi^*=0$. Top panel: 2D (magenta) and 3D (green) stability indexes  with zoom of {\it vco}. The black vertical line in the zoom panel locates the {\it vco}. Low panel: semi-major axis $a$  and eccentricity $e$  (horizontally stable (red / blue).}
 \label{1}
 \end{figure}

\begin{figure}
  \centering   
\includegraphics[width=\textwidth]{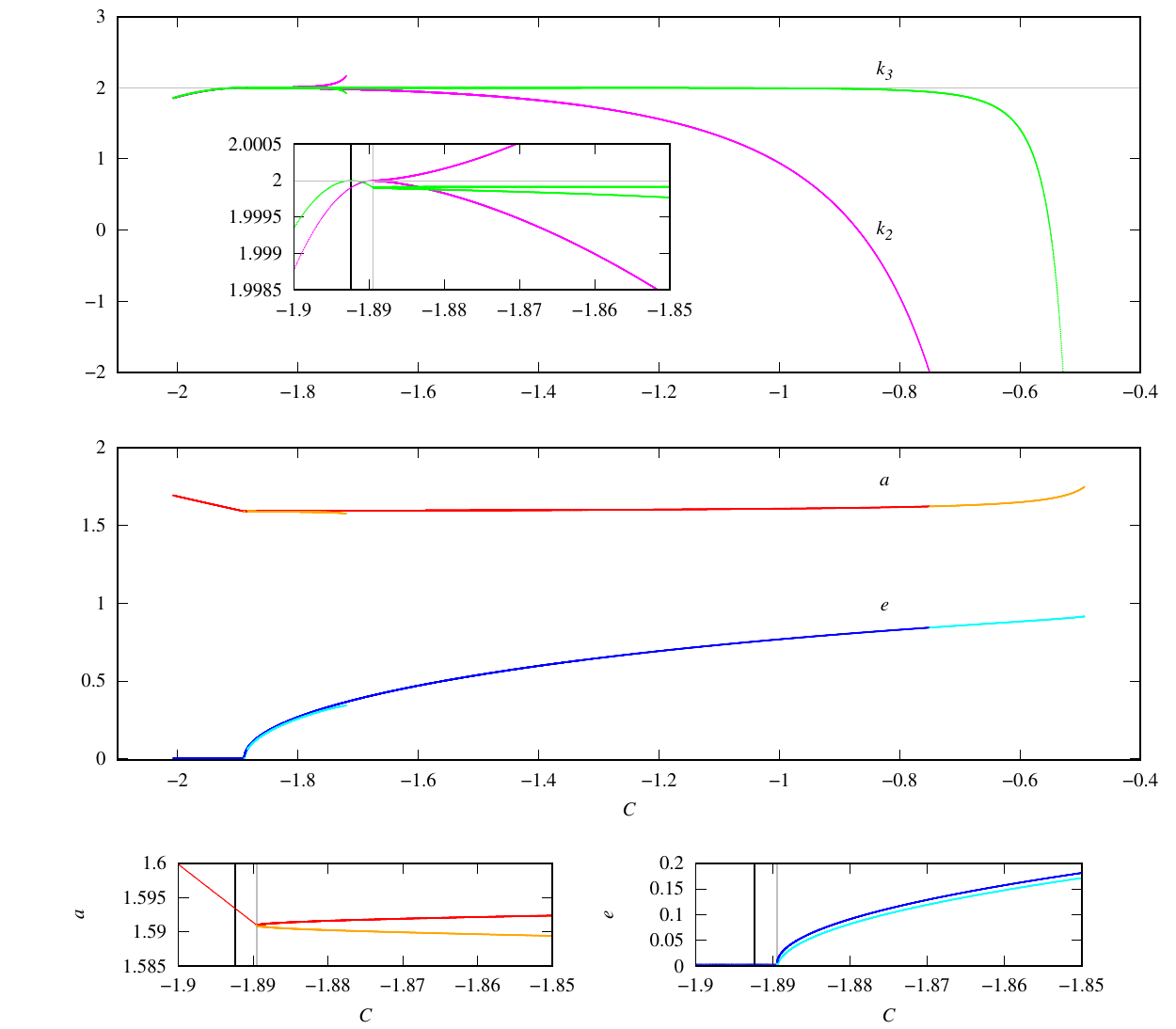}
 \caption{The nearly circular SPO family and its bifurcations into a stable family (corresponding to the resonance center  $\phi^*=180^\circ$)  and an unstable family (corresponding to $\phi^*=0$). Top panel: 2D (magenta) and 3D (green) stability indexes  with zoom of {\it vco} and 2D bifurcation. Middle panel: semi-major axis $a$  and eccentricity $e$  (horizontally stable (red / blue) and unstable (orange / cyan)). Low panel: zoom of bifurcation region. Vertical lines in the zoom panels locate the {\it vco} (black line) and the 2D bifurcation of the circular family into a stable / unstable pair (gray line).}
\label{2}
\end{figure}

We show how the planar  families  of symmetric POs (SPOs) evolve with the Jacobi constant, $C$, in  Fig.~\ref{1} (resonant mode with $\phi^*=0$) and Fig.~\ref{2} (circular family  and resonant mode with $\phi^*=180^\circ$ ).  

The eccentric family  (named $R_{Ib}$ in \citet{KotoulasVoyatzis2020b}) is horizontally stable  except for a barely visible small segment at the start near collision with the planet.  This family corresponds to the resonance center $\phi^*=0$.   There is a {\it vco} at $C=-1.5649$ and 3D quasiperiodic orbits around the family exist when $C>-1.5649$. The family occurs above the pericentric collision line with the planet  \citep{MoraisNamouni2016JCAM}, and ends by collision with the star. We call this the high eccentricity resonant mode.

The circular family, when computed with multiplicity 3, is vertically stable with a single {\it vco} at $C=-1.8924$. At $C=-1.8895$ (where the ratio of mean motions is very close to 1/2) the horizontal stability index  on the circular family reaches the value $k_2=2$ and a  bifurcation into a pair of stable / unstable POs  occurs.  The unstable branch corresponds to $\phi^*=0$ (family $R_{Ia}$ in \citet{KotoulasVoyatzis2020b}). The  stable branch  corresponds to the resonant mode  $\phi^*=180^\circ$ (family $R_{IIa}$ in \citet{KotoulasVoyatzis2020b}) and is horizontally  stable from the bifurcation when $e\approx0$ up to $C=-0.7505$ (when $e\approx 0.8$) near collision with the planet which occurs midway between apocentre and pericentre
 \citep{MoraisNamouni2016JCAM}.  When $C>-0.7505$ the family is horizontally unstable and when $C>-0.5286$ it is also vertically unstable. At $C=-0.7505$ and $C=-0.5286$ the stability indexes are $k_2=-2$ and $k_3=-2$, respectively. Associated possible bifurcation into 2D and 3D families with multiplity 6 are very difficult to confirm due the the strong chaos that exists at these locations.  Note that the stable family which exists up to eccentricity nearly 1 at Neptune to Sun mass ratio  (named $R_{IIb}$ in \citet{KotoulasVoyatzis2020b}) does not occur at Jupiter to Sun mass ratio.

\subsection{Periodic orbits in 3D}

As shown by \citet{NamouniMorais2020} in the 3-dimensional case the relevant resonant angles are  $\phi_k=2\lambda-\lambda_p-\Omega-k\omega$, where $\omega$ is the argument of pericenter and $k$ is an odd integer for the 1/2 resonance. When $k=1$ we have the planar prograde angle $\phi_1=\phi=2\lambda-\lambda_p-\Omega-\omega$ and when $k=3$ we have the planar retrograde angle $\phi_3=\phi^*=2\lambda-\lambda_p-\Omega-3\omega$, respectively. Recall that when using the modified (retrograde) angles from \citet{MoraisNamouni2013CMDA} we have  $\phi=2\lambda^*+\lambda^*_p-\varpi^*+2\Omega$ and $\phi^*=2\lambda^*+\lambda^*_p-3\varpi^*$. These modified angles are part of a canonical set of variables adequate to study retrograde resonance. Particularly, they give us information about retrograde resonance structure: $\phi$ is associated with a mixed type 3rd order resonance \citep{NamouniMorais2015} while $\phi^*$ is associated with an eccentricity 3rd order resonance \citep{MoraisNamouni2013CMDA}.

The planar SPOs from the previous section may be continued into 3-dimensional SPOs  starting at the {\it vcos} which occur at:
 $C=-1.8924$ (circular family); $C=-1.5649$  and $e=0.4966$ (resonant family corresponding to $\phi^*=0$).  In general, SPOs correspond to the resonant centers 0 or $180^\circ$. The 1/2 resonance may also have  asymmetric periodic orbits (APOs) which correspond to  resonant centers different from  0 and $180^\circ$ \citep{Voyatzisetal2018}, \citep{NamouniMorais2020}.

\begin{figure}
  \centering
    \includegraphics[width=\textwidth]{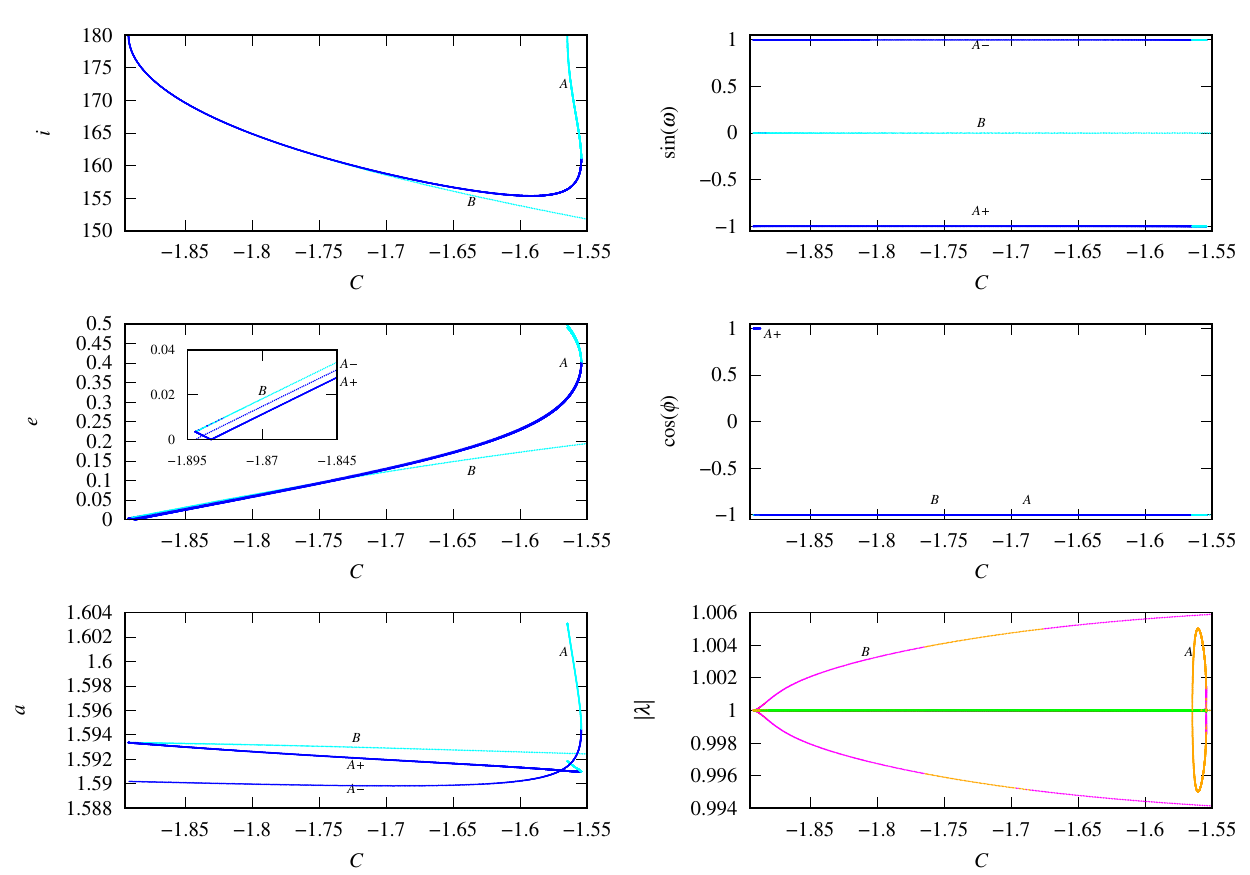}
 \caption{Families of 3D  symmetric periodic orbits bifurcating from the {\it vcos}. Left panels: inclination $i$; eccentricity $e$ with zoom; semi-major axis $a$. Right panels: cosine of argument of pericenter $\omega$; cosine of prograde resonant angle $\phi=\phi_1$; eigenvalues amplitudes $|\lambda|$. The families are coloured blue (cyan) when stable (unstable).  There are 2 families bifurcating from the {\it vco} on the circular planar family. One has type (a) symmetry, is initially stable (complex conjugate eigenvalues with unit amplitude), becomes unstable but nearly critical  at $C=-1.5545$ (one pair of real eigenvalues with unit amplitude) and ends at the {\it vco} on the resonant eccentric planar family. The other has type (B) symmetry and is unstable but initially nearly critical (one pair of real reciprocal eigenvalues with amplitudes close to unit).  The familes are labeled $b$ (unstable type (B)) and $a$ ( type (a)). The branches labeled $a+$ and $a-$ refer to distinct intersections of the type (a) family with the surface of section which are   obtained by continuation from the {\it vcos} at $C=-1.8924$ and $C=-1.5649$, respectively.
 }
 \label{3}
 \end{figure}

Figure \ref{3} shows the 3D  SPOs bifurcating from the {\it vcos} on the planar families.  There are 2 symmetric families bifurcating from  the {\it vco} at $C=-1.8924$ on the circular  family. One  has type (a) or  F symmetry and is initially stable  but changes stability at $C=-1.5545$  (clearly seen in Fig.~\ref{3} bottom right panel) which corresponds to the maximum value of the Jacobi constant. The other has type (B)  or G symmetry and is always unstable but initially nearly critical  due to the proximity to the {\it vco}.  Kotoulas \& Voyatzis (2020b) identify a similar bifurcation when  using a   Neptune to Sun  mass ratio, into type (a) and type (B) SPOs which they classify as  R01  and  R0G families.
Continuation of the type (a) family  shows that  it connects to the {\it vco}   at $C=-1.5649$ on the 2D family corresponding to the resonant mode $\phi^*=0$,   similarly to the case with Neptune to Sun mass ratio (Kotoulas \& Voyatzis 2020 b). However, note that in the latter case   no  change of stability along the family was detected.

\begin{figure*}
\includegraphics[width=6cm]{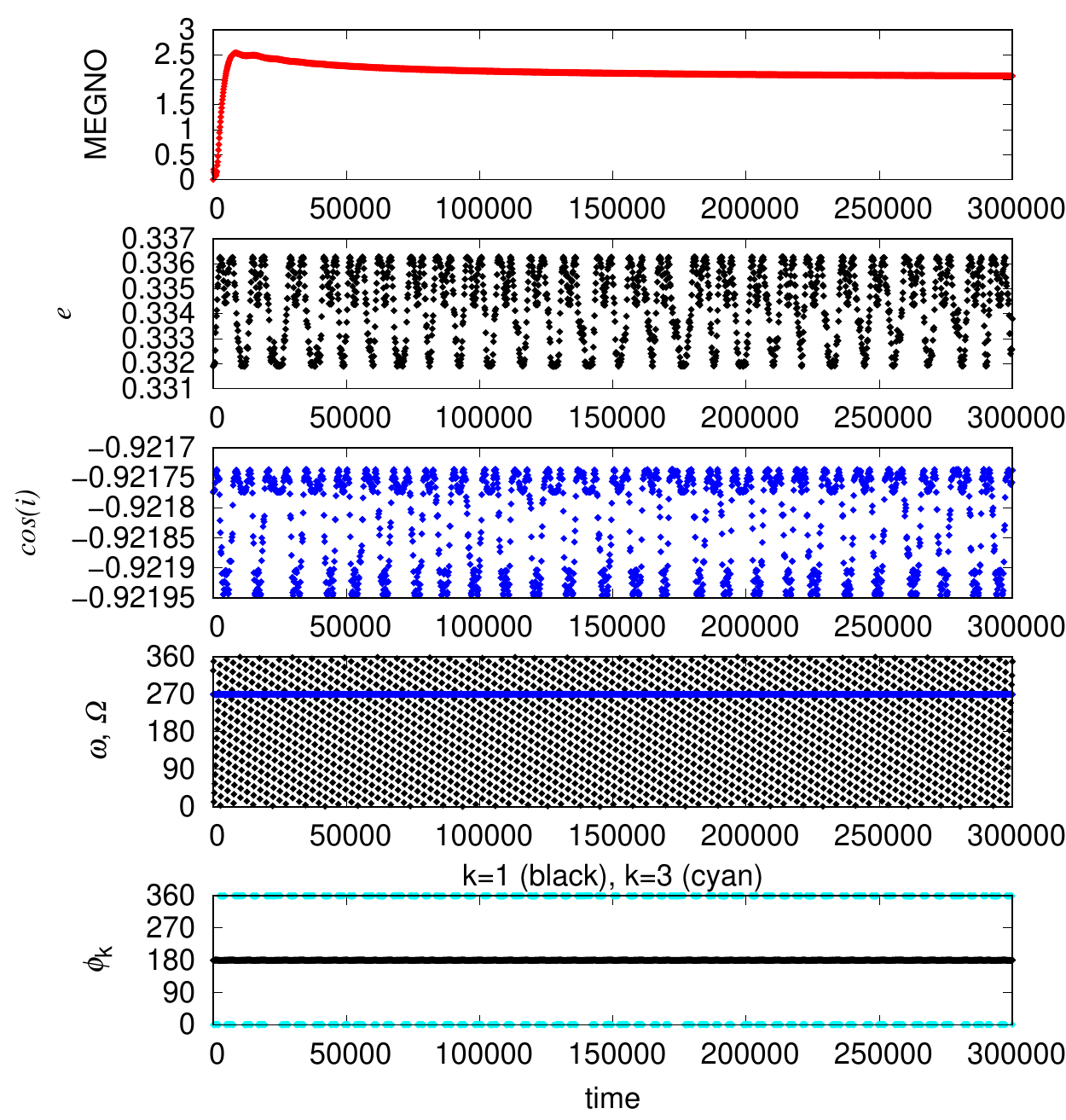}
\includegraphics[width=6cm]{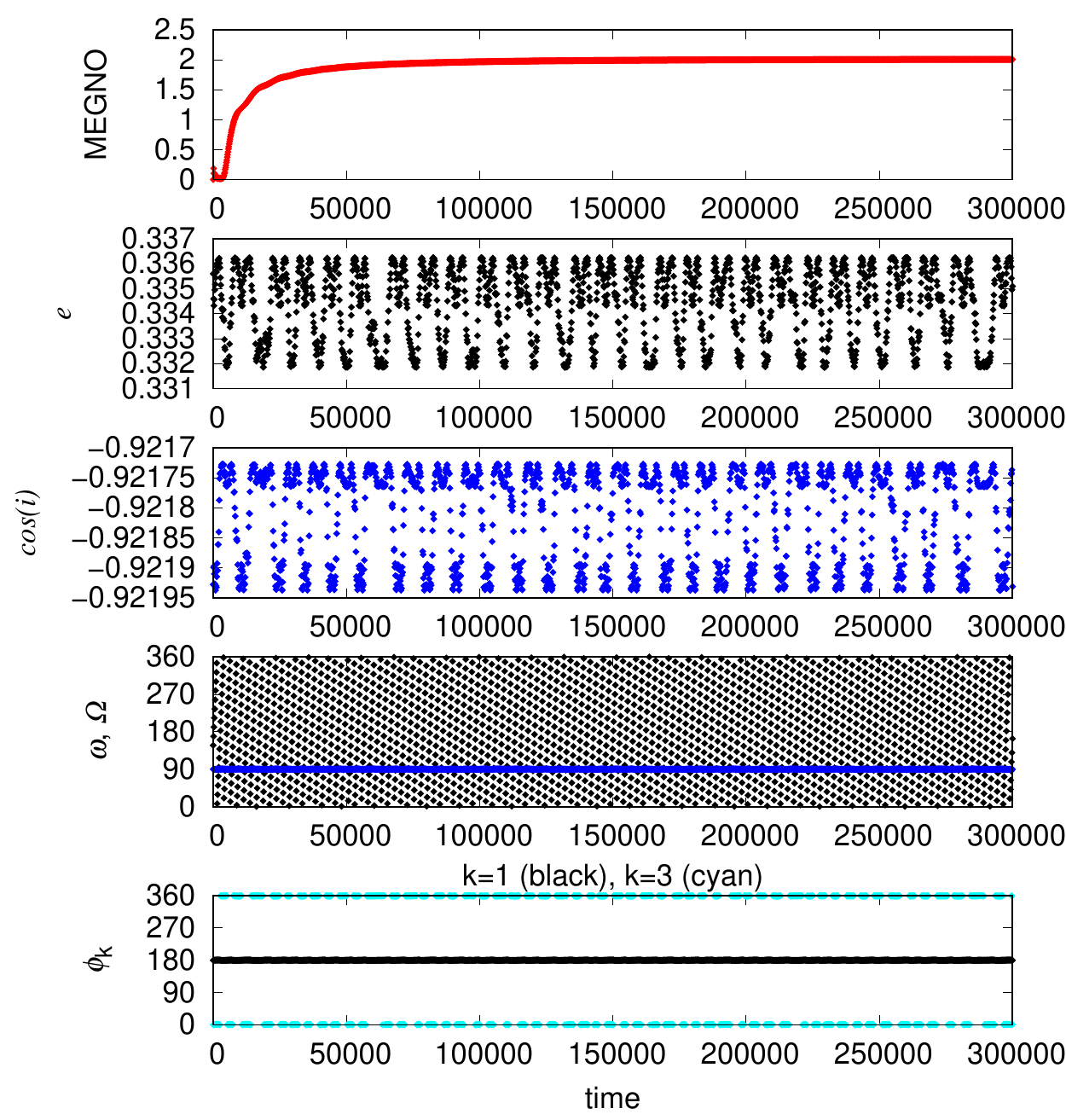} 
\includegraphics[width=6cm]{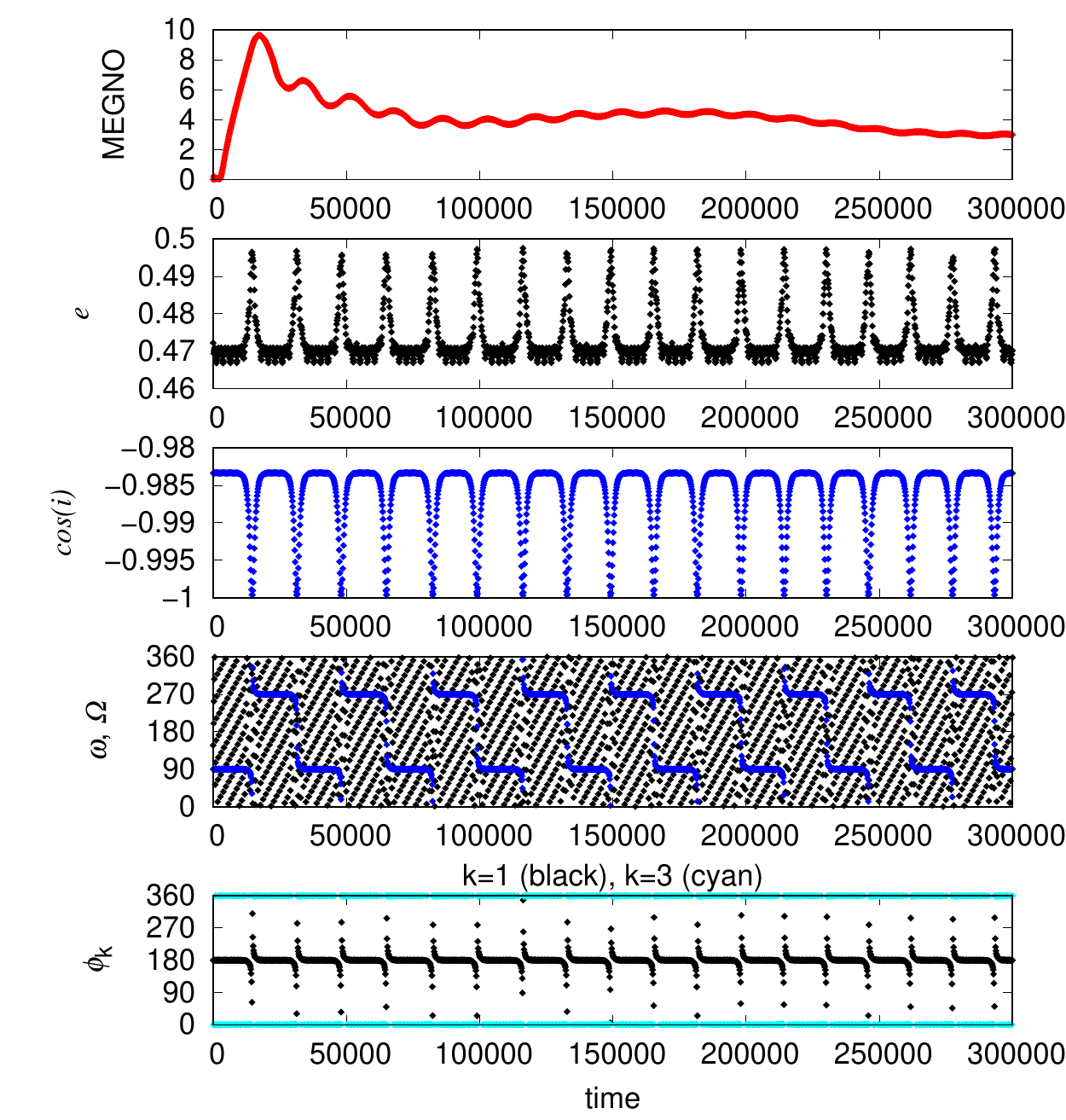}\includegraphics[width=6cm]{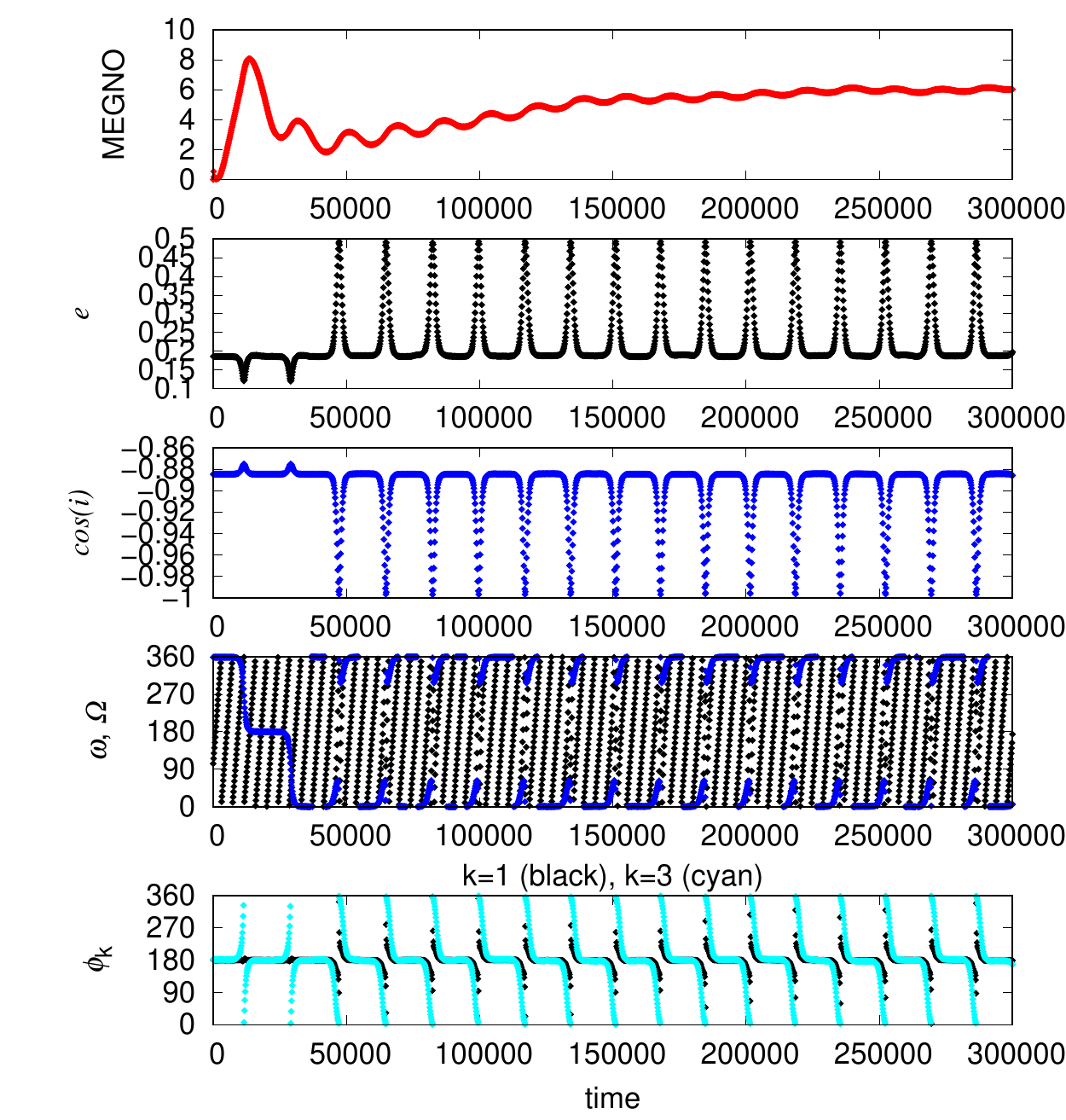}
 \caption{Top panels: Type (a) SPOs at $C=-1.5600$ (stable branches  originating from bifurcations from the {\it vcos} on the circular family (left) and  resonant family corresponding to $\phi^*=0$  (right)). Low left panel: Type (a) SPO at $C=-1.5600$ (unstable nearly critical branch). Low right panel: Type (B) SPO at $C=-1.5600$. 
 From top to bottom panels:  MEGNO; eccentricity, cosine of inclination, argument of pericenter $\omega$ (blue) and longitude of ascending node $\Omega$ (black); resonant angles $\phi=\phi_1$ (black) and $\phi^*=\phi_3$ (cyan). }
 \label{4}
 \end{figure*}

The type (a) SPO family which bifurcates from the nearly circular family, initially corresponds to the fixed point of the resonance Hamiltonian with $\phi=0$, $\omega=270^\circ$ i.e.\ $\phi^*=180^\circ$. The eccentricity along the family decreases until the value zero is reached and then increases again (Fig.~\ref{3}, middle left). At this point  there is a shift to the fixed point  with $\phi=180^\circ$, $\omega=270^\circ$ i.e.\ $\phi^*=0$   (Fig.~\ref{3}, top \& middle right). This change of topology  is reproduced in our Hamiltonian model for  3D retrograde resonance  described in the next section. 
Continuation  from the {\it vco} at $C=-1.5649$ retrieves the same family but at a different intersection with the surface of section which corresponds to the fixed point of the resonance Hamiltonian   with $\phi=180^\circ$, $\omega=90^\circ$ i.e.\ $\phi^*=0$   (Fig.~\ref{3}, top \& middle right). In Fig~\ref{3}, the labels $a+$ and $a-$ are used to distinguish the 2 intersections of the (a) type family with the surface of section. Examples of POs on the sections $a+$ and $a-$ are shown in Fig.~\ref{4} (top left and right panels). On the  unstable nearly critical  branch bifurcating from the {\it vco} at $C=-1.5649$ on the resonant family,  $\phi^*=0$ with $\omega$ circulating on  several thousand years timescale around the Kozai centers at $\omega=0,180^\circ$  (Fig.~\ref{4}, bottom left  panel). Near the transition to stability at $C=-1.5545$,  MEGNO  increases linearly with time at a slow rate  (Fig.~\ref{4}, bottom  left panel).   MEGNO converges to 2  on the family's stable branch since motion is regular (Fig.~\ref{4}, top panels). 

The type (B) SPO  corresponds to the  unstable fixed points $\phi=180^\circ$,  $\omega=0,180^\circ$ (Fig.~\ref{3}, top \& middle right).  After several thousand years, there is alternating libration and circulation around $\omega=90^\circ,270^\circ$ which indicates  the presence of  Kozai separatrices (Fig.~\ref{4}, bottom right panel) with typical coupled eccentricity and inclination oscillations. Again MEGNO  increases linearly with time indicating chaotic diffusion .

\begin{figure*}
\includegraphics[width=4cm]{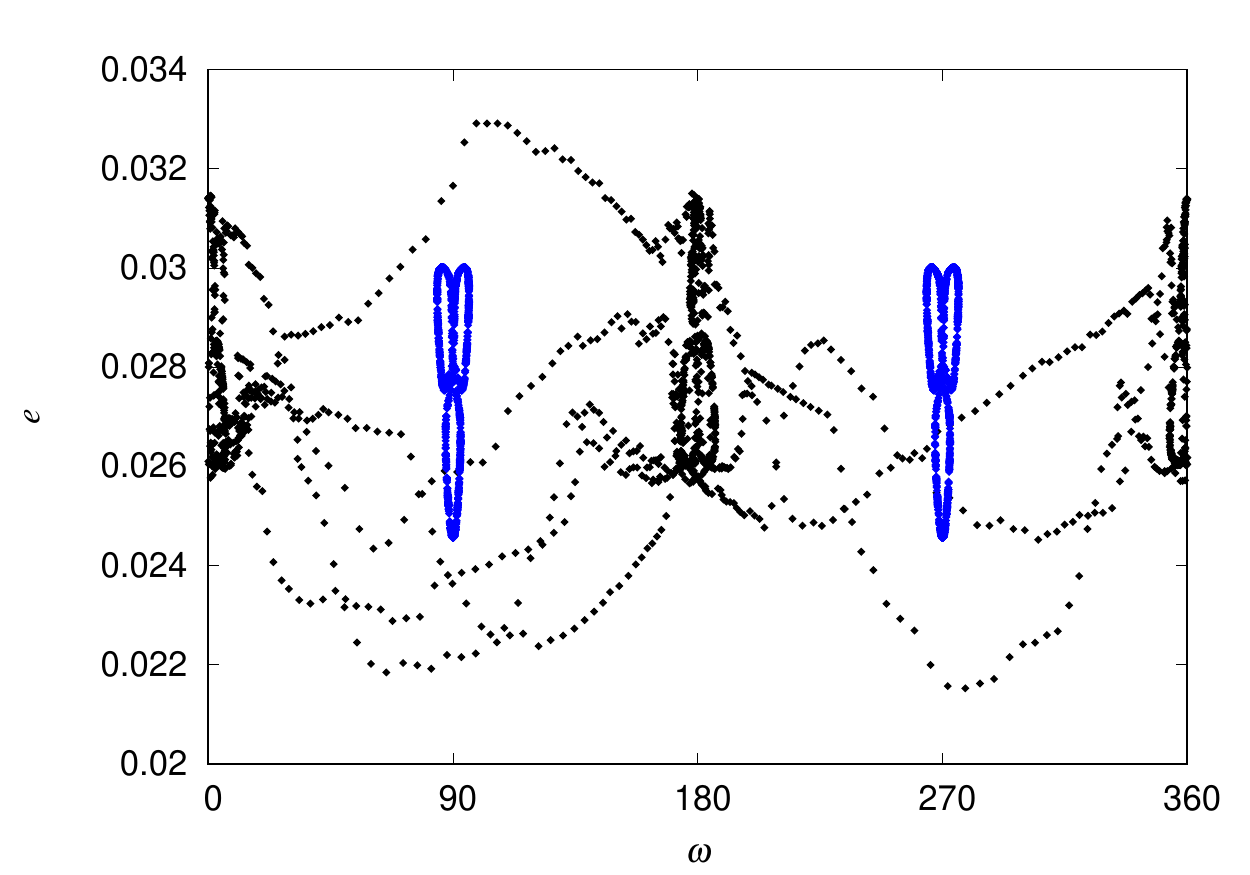}
\includegraphics[width=4cm]{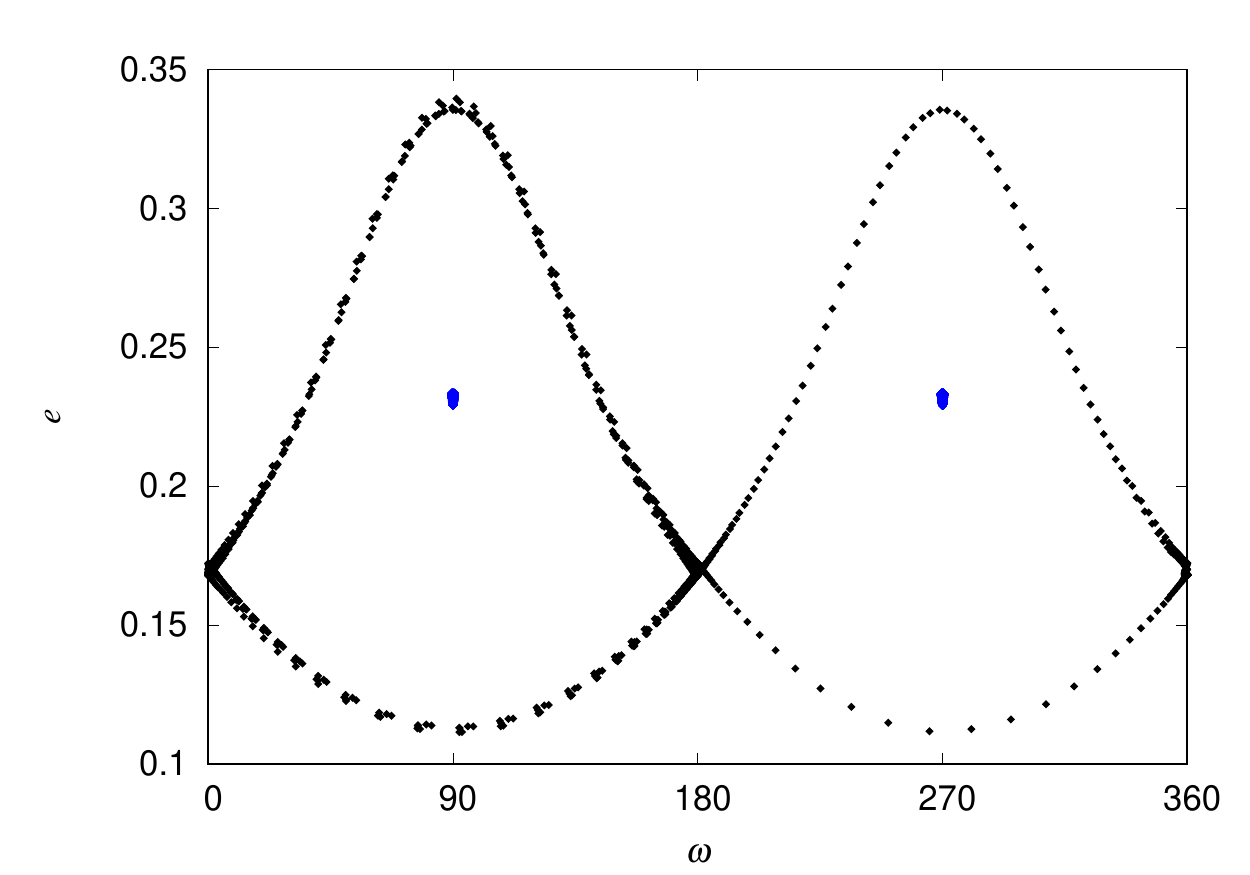} 
\includegraphics[width=4cm]{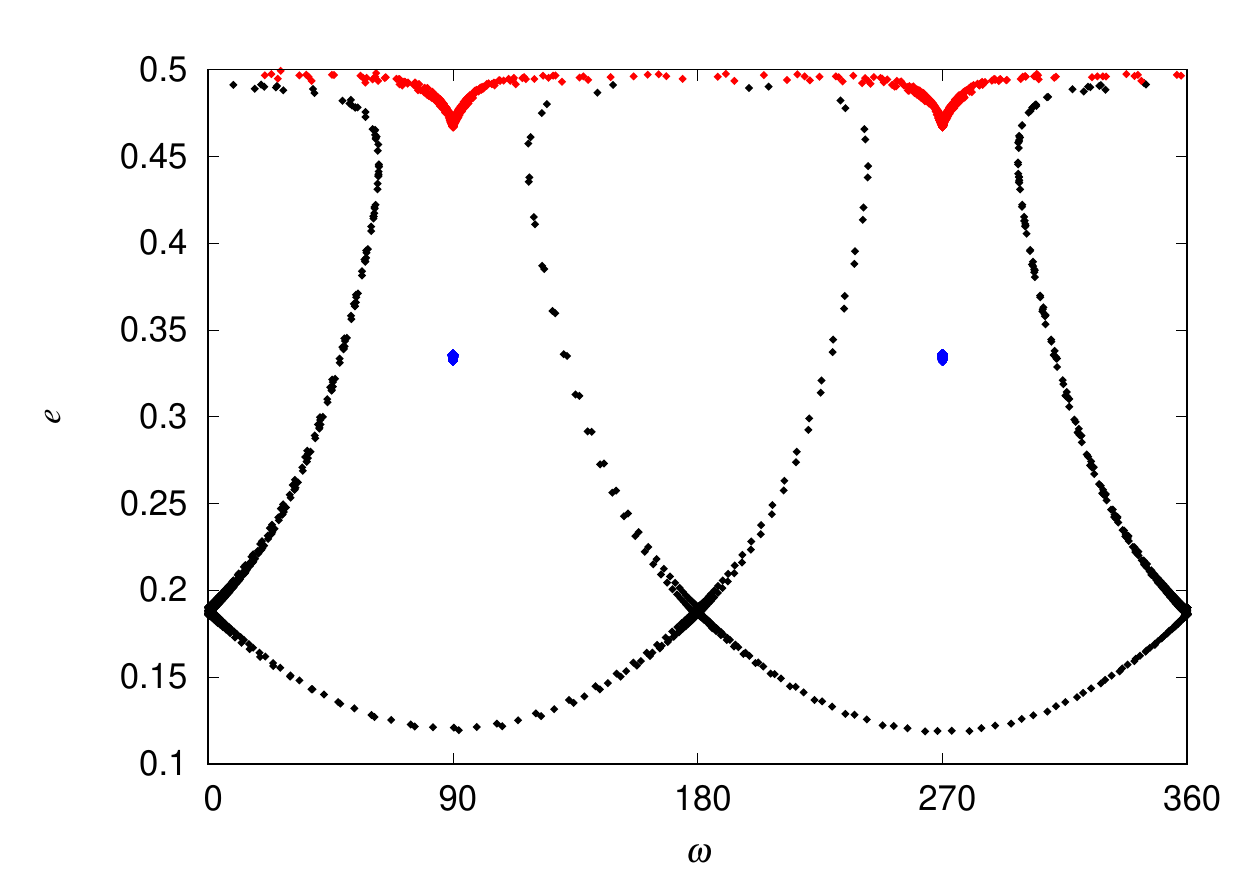}
 \caption{Kozai diagrams for type (B) SPO (black), stable type (a) SPOs (blue) and nearly critical type (a) SPO (red) at  $C=-1.8500$ (left), $C=-1.6000$ (middle) and $C=-1.5600$ (right). Evolution of the orbits over $3\times 10^5$ planet's periods.}
 \label{5}
 \end{figure*}

The connection between the 2 families becomes clearer when plotting the orbits' evolution  in Kozai-type $(e,\omega)$ diagrams (Fig.~\ref{5}).
While the stable type (a) SPO family corresponds to the Kozai centers at $\omega=90^\circ,270^\circ$ (blue), the unstable but nearly critical type (B) SPO family evolves towards Kozai separatrices around these centers (black). 
At $C=-1.8500$  the unstable nearly critical type (B) SPO  overlaps with the  stable centers corresponding to the type (a) SPO family while  at $C=-1.6000$ and $C=-1.5600$ the 2 families are well separated. At $C=-1.5600$ the unstable nearly critical type (B) SPO corresponding to $\phi=180^\circ$ connects with the nearly critical branch on the type (a)  SPO corresponding to $\phi^*=0$ and circulation around the Kozai centers at $\omega=0,180^\circ$ (red).

\begin{figure}
  \centering
    \includegraphics[width=\textwidth]{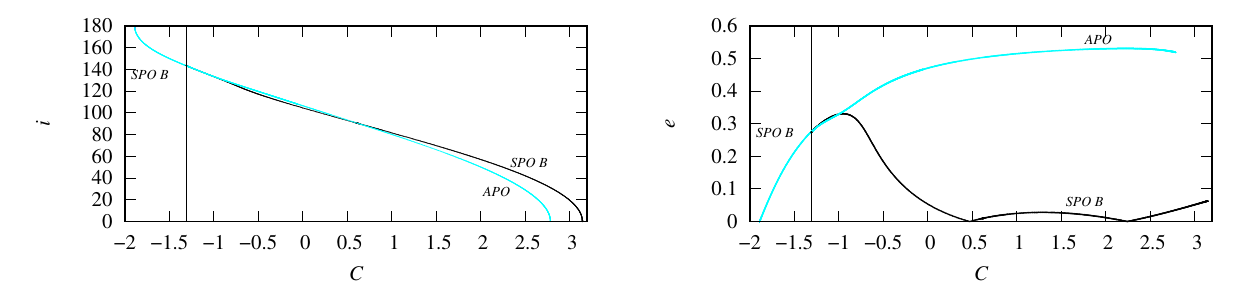}
 \caption{The bifurcation of the type (B) symmetric family ($SPO\, B$) into an single unstable asymmetric family ($APO$) and a double unstable symmetric family ($SPO\,B$). Left panel: inclination $i$. Right panel: eccentricity $e$. The black vertical line indicates the bifurcation at $C\approx-1.308$. The colours cyan/black indicate if the family is single unstable (one $\lambda_i$ larger than 1) or double unstable (two $\lambda_i$ larger than 1).}
 \label{6}
 \end{figure}

The type (a) SPO family exists in the range $-1.8924<C<-1.5545$. When $C>-1.5545$ only the  type (B) SPO family persists.   Instability, measured by the maximum eigenvalues' amplitudes, increases with $C$ (Fig.~\ref{3} bottom right). At $C\approx -1.308$ the family divides into 2 branches of POs with different properties which could be continued to $i=0$ (Fig.~\ref{6}). One branch corresponds to the SPO type (B) family  which becomes double unstable (two eigenvalues' amplitudes, $|\lambda|$, larger than 1) at the bifurcation point. This branch connects with the prograde $A_2$ family from \citet{Kotoulas&Hadji2002}.  The 2nd branch is single unstable (one eigenvalue amplitude, $|\lambda|$, larger than 1) asymmetric and  corresponds to the continuation of family $A_{22}$  identified in the prograde problem by \citet{Voyatzisetal2018}.

\section{Resonance capture in 2D and 3D}

We simulate resonant capture following the setup in \citet{NamouniMorais2015,NamouniMorais2017}. Stoke's drag causes the test particle's semimajor axis to drift towards the resonance following an exponential decay law $a=a_0 \exp^{-t/\tau}$. We ensure adiabatic capture by choosing a long characteristic decay time  $\tau=10^7$ planet's periods.

\begin{figure*}
\includegraphics[width=3cm]{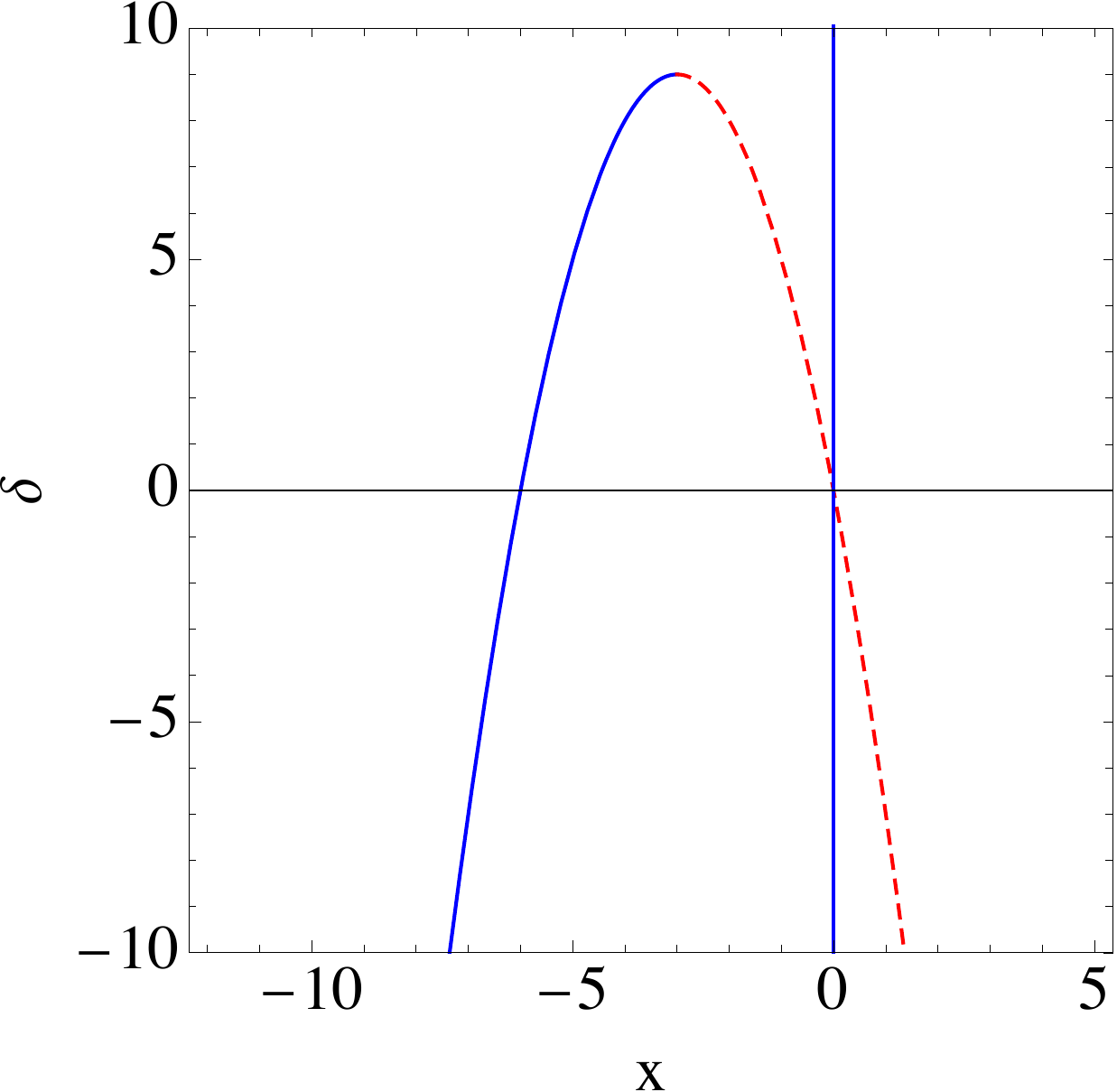}
\includegraphics[width=3cm]{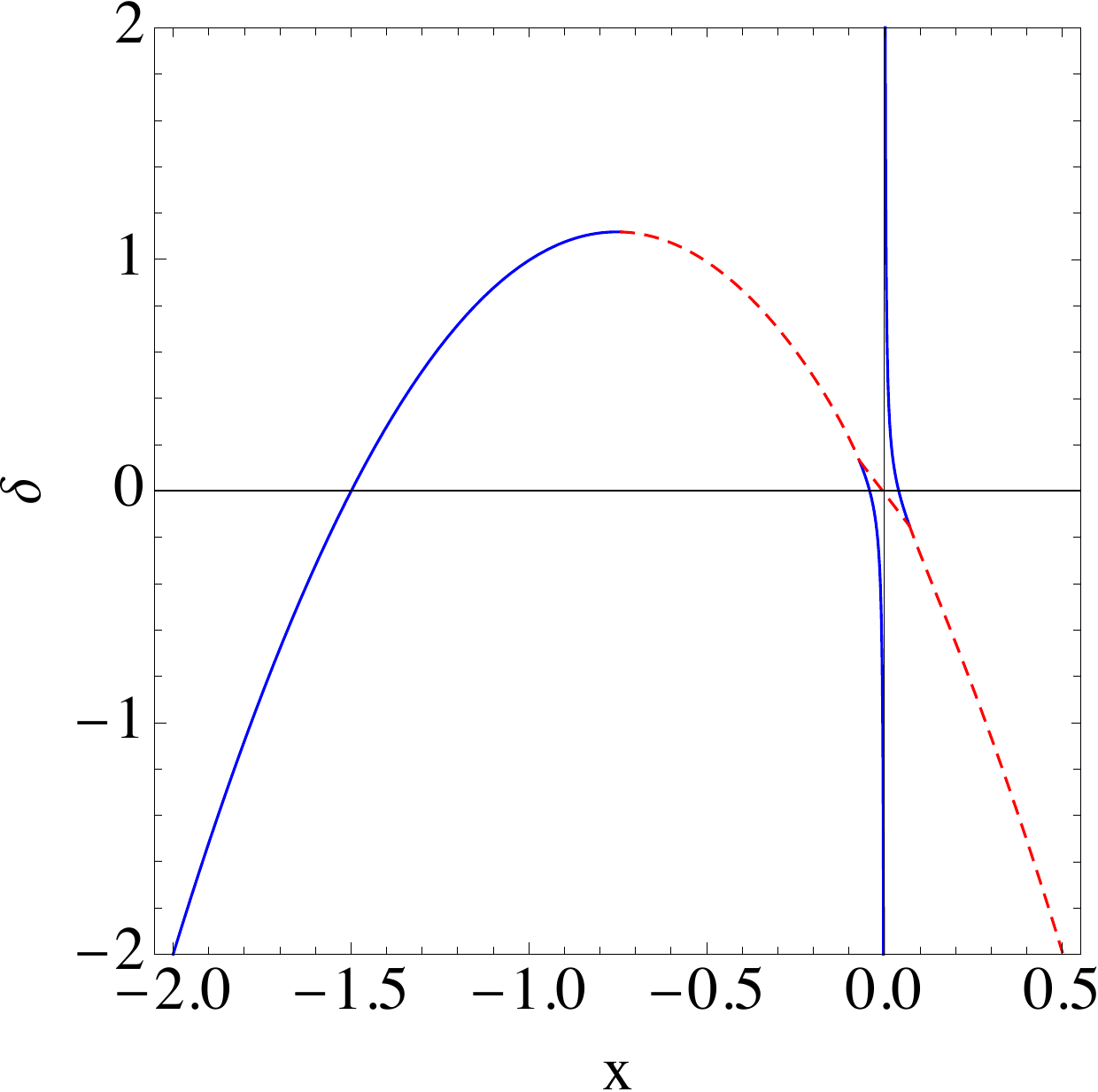}
\includegraphics[width=3cm]{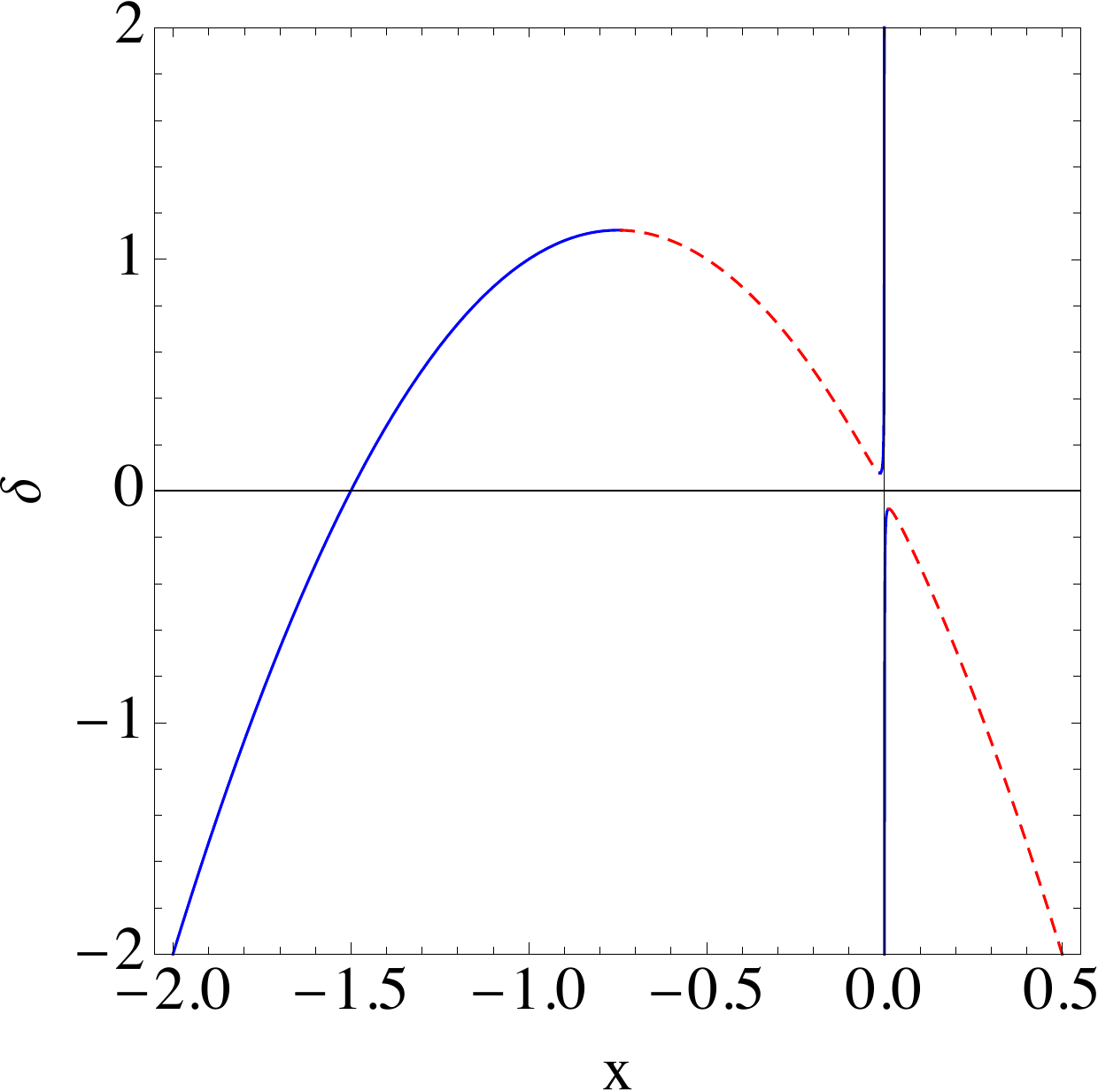}
\includegraphics[width=3cm]{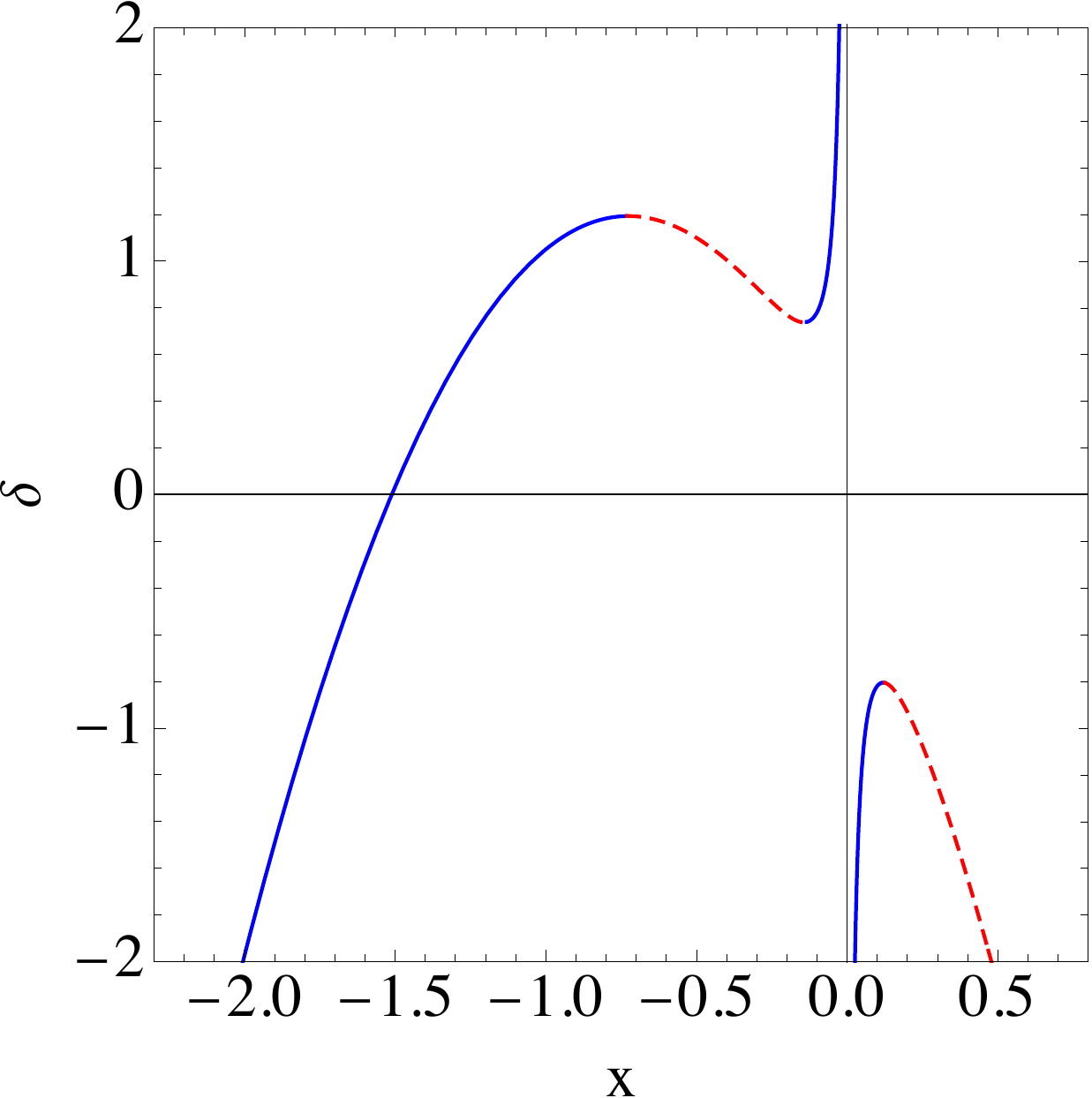} 
 \caption{Bifurcation diagrams of the Hamiltonians for retrograde 3rd order resonance:  2D case from \citet{MoraisGiuppone2012} with $H=\delta (x^2 + y^2 )/2 + (x^2 + y^2 ) ^2/4 -2 x (x^2-3 y^2)$ and $x\propto e^3 \cos{\phi_3}$ (left); 3D case from \citet{NamouniMorais2015} with $H=(\delta+2 x) (x^2 + y^2 ) + (x^2 + y^2 ) ^2 + x T$ and $x\propto e \cos^2(i/2) \cos{\phi^*}$, $T=-0.01$ (middle left) $T=0.001$ (middle right) and $T=0.1$ (right). The parameter $\delta \propto n_p-2\,n$  decreases as the particle drifts towards resonance, while $T$ is a proxy for $(\pi-i)^2-2\,e^2$. Stable (unstable)  fixed points are shown in solid blue (dashed red). The vertical line locates $x=0$.}
 \label{7}
 \end{figure*}

In the planar problem, outer  orbits slowly approaching the planet follow the nearly circular non-resonant family which bifurcates into a resonant  SPO corresponding to $\phi^*=180^\circ$.  However, capture in the 1/2 retrograde resonance never occurs for planar circular orbits. As shown by \citep{MoraisGiuppone2012,MoraisNamouni2013CMDA}, the  resonant term is $e^3 \cos{\phi^*}$.  The equilibrium points of an Hamiltonian model for this planar 3rd order resonance valid at small to moderate eccentricities \citep{ssdbook} are shown in the bifurcation diagram $(x,\delta)$, where $x\propto e^3 \cos{\phi^*}$ and $\delta\propto n_p-2\,n$ (Fig.~\ref{7} left). When exact resonance ($\delta=0$) is reached,  the unstable fixed point is at $e=0$.  Therefore, the final outcome  will always be $\phi^*$ circulating outside the resonant separatrix.

3D  nearly circular outer orbits  drifting towards the planet are 1st captured in the resonant mode $\phi=180^\circ$. In \citet{NamouniMorais2015} we  explained this with  an Hamiltonian model of the mixed resonance, $e \cos^2(i/2) \cos{\phi}$,  which has an additional parameter, $T$, a proxy for $(\pi-i)^2-2\,e^2$.  In Fig.~\ref{7} we show the corresponding bifurcation diagrams $(x,\delta)$ where $x\propto e \cos^2(i/2) \cos{\phi}$, for different values of $T$.  Before resonance passage ($\delta>0$) when $T>0$ there are 3 apocentric ($x<0$, i.e.\ corresponding to 
$\phi=180^\circ$) equilibrium points: 2 stable and 1 unstable, but as resonance ($\delta=0$) is approached only the higher eccentricity stable apocentric equilibrium (branch on the left) remains (Fig.~\ref{7} right), similar to a 1st order resonance. Therefore, capture in $\phi=180^\circ$ occurs with $100\%$ probability. Indeed, we confirmed numerically that this occurs when  $i\le174.7^\circ$.  As the inclination increases, the parameter $T$ decreases and the unstable apocentric equilibrium moves closer to $x=0$ (Fig.~\ref{7} middle right).  When $T=0$  the  topology of the 3rd order mixed resonance  is similar to that of the eccentricity resonance (Fig.~\ref{7} left):  the unstable fixed point is  exactly at $x=0$ when $\delta=0$. When the inclination approaches $180^\circ$, $T$ becomes slightly negative (Fig.~\ref{7} middle left) and when $\delta>0$ the stable equilibrium at small eccentricity has $x>0$ (pericentric i.e.\ corresponding to $\phi=0$) in agreement with the behaviour at the start of the type (A) SPO family described in the previous section. Since the unstable equilibrium is always near $x=0$, capture in resonance never occurs for nearly circular orbits with $i>174.7^\circ$.

\begin{figure*}
\includegraphics[width=6cm]{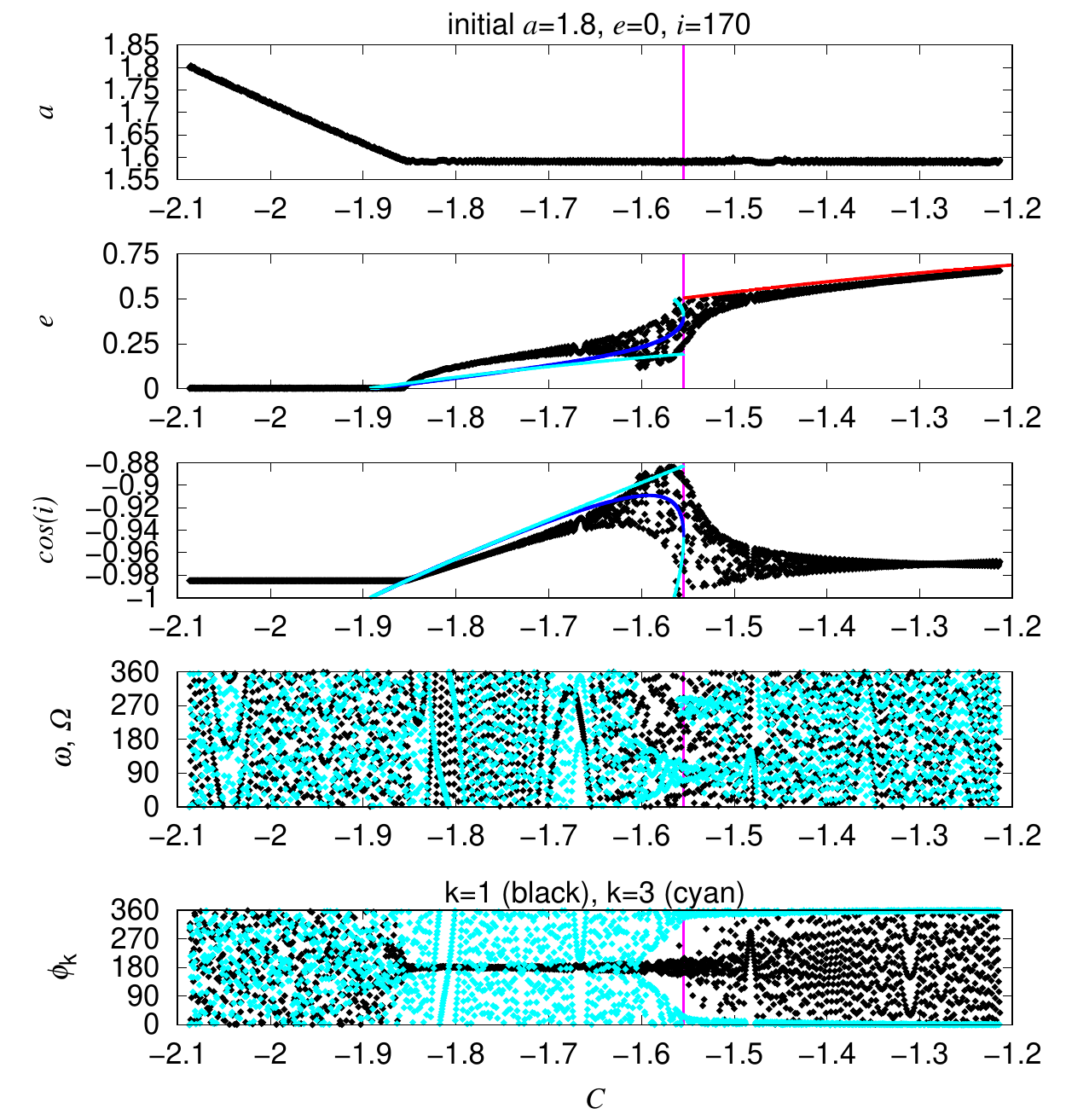}
\includegraphics[width=6cm]{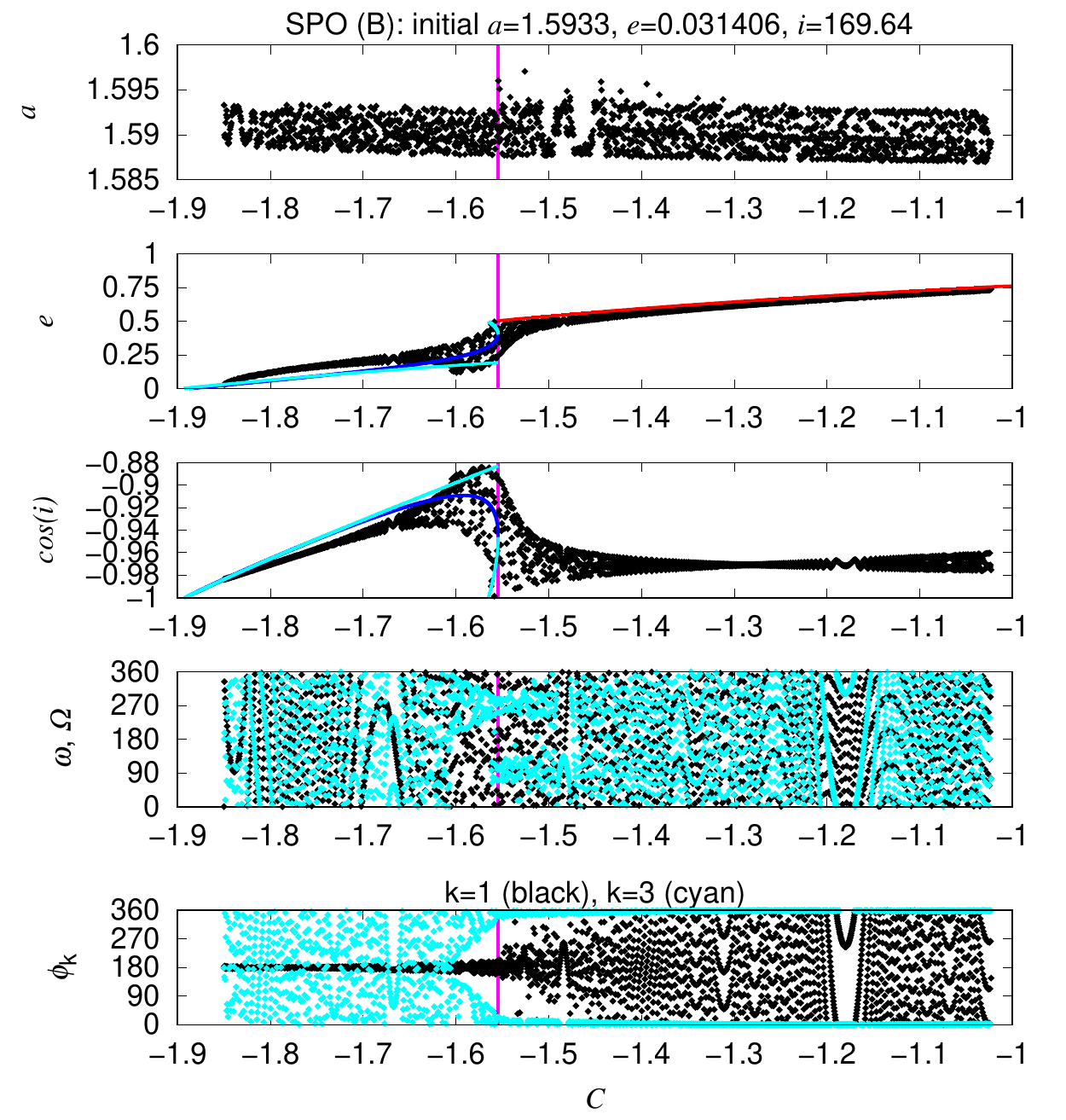}
\includegraphics[width=6cm]{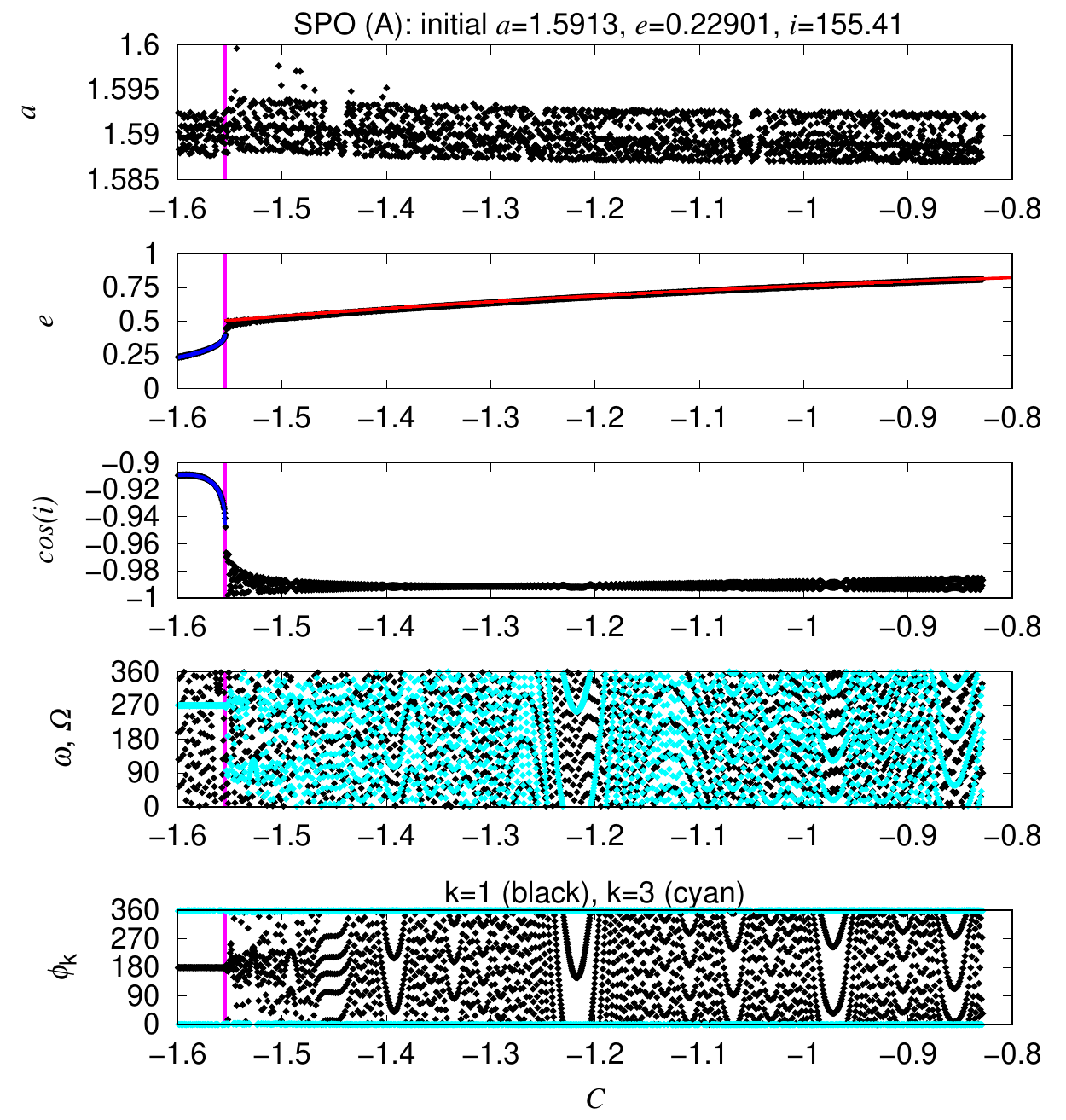}
\includegraphics[width=6cm]{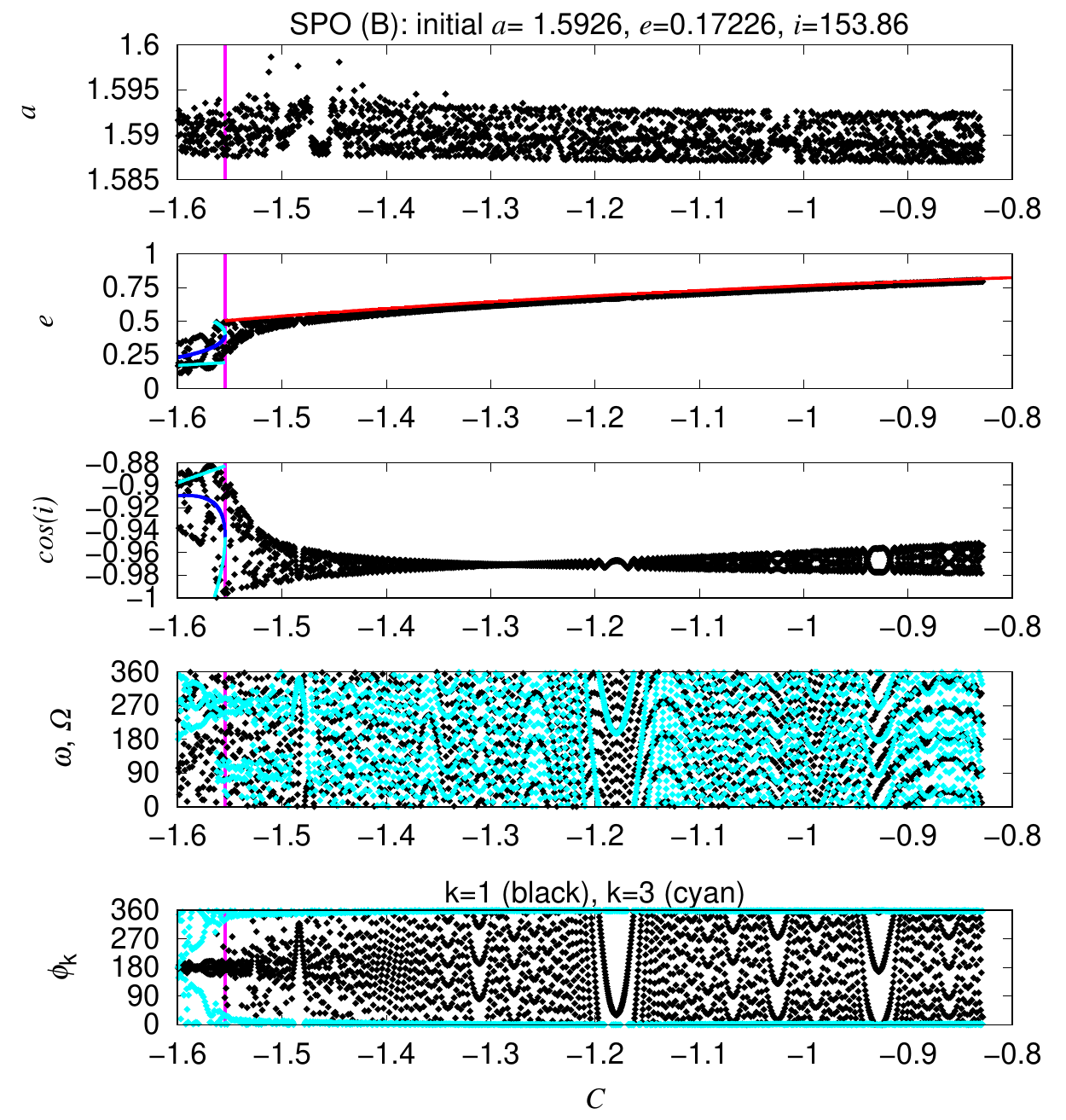}
 \caption{Evolution of orbits  subject to Stoke's drag  (semimajor axis exponential decay law with characteristic time $\tau=10^7$ planet's periods).
Top left: initial $a=1.8$, $e=0$, $i=170^\circ$. Top right: initially at the SPO  type (A) or type (B) family with $C=-1.8500$. Bottom left / right: initially on the type (A) / type (B) family with $C=-1.6000$.
From top to bottom panels:  semi-major axis; eccentricity, cosine of inclination, argument of pericenter $\omega$ (blue) and longitude of ascending node $\Omega$ (black); resonant angles $\phi=\phi_1$ (black) and $\phi^*=\phi_3$ (cyan).}
\label{8}
\end{figure*}

A typical example of 3D capture with  $a_0=1.8$, $e=0$, $i=170^\circ$ is shown in Fig.~\ref{8} (top left).  At $C\approx -1.856$ there is capture in $\phi=180^\circ$  with an increase of eccentricity and decrease of inclination characteristic of a mixed type $e \cos^2(i/2) \cos{\phi}$ resonance. At $C\approx-1.604$ there is  Kozai libration around $\omega=90^\circ$ with  large coupled oscillations in eccentricity and inclination which lead to capture in $\phi^*=0$  at $C\approx-1.562$  when $e>0.4$, with an associated increase of eccentricity characteristic of an eccentricity $e^3 \cos{\phi^*}$ resonance. This 3-stage capture mechanism was first observed in the numerical simulations from \citet{NamouniMorais2015} and occurs for nearly circular orbits with $175^\circ \gtrsim i \gtrsim 155^\circ$.

We follow  initial conditions on the SPO   families to understand their role in resonant capture. At $C=-1.8500$ the evolution of initial conditions on the type (A) and type (B) families are indistinguishable, due to their proximity (Fig.~\ref{5}, left). The presence of dissipation implies that there is a drift from the stable resonance  centers associated with the type (A) family and the test particle follows the typical capture path  as if  it was started on the  unstable nearly critical type (B) family (Fig.~\ref{8} top right).  At $C=-1.6000$  the 2 familes are now separated hence  the test particle remains on the type (A) SPO family which corresponds to the fixed point $\phi=180^\circ$, $\omega=270^\circ$ and $\phi^*=0$ (Fig.~\ref{8} bottom left),  with eccentricity  and inclination increasing as $C$ increases due to dissipation. At $C=-1.5545$, a saddle-node bifurcation occurs on the type (A) SPO family (Fig.~\ref{3}) but the eccentricity is large enough for capture in a 3D quasiperiodic orbit around the 2D resonant mode $\phi^*=0$.
The initial condition on the type (B) family at $C=-1.6000$ (Fig.~\ref{8} bottom right) evolves towards the Kozai separatrix around the centers $\omega=90^\circ,270^\circ$ on the  type (A)  family, with typical coupled oscillations of eccentricity and inclination. The eccentricity of these Kozai centers increases as the semi-major axis decreases. Once the saddle-node bifurcation on the  type (A) SPO family occurs at $C=-1.5545$, there is capture  in a 3D quasiperiodic orbit around the 2D resonant mode $\phi^*=0$ which is vertically stable when $C>-1.5649$. 
 
\begin{figure*}
\includegraphics[width=6.0cm]{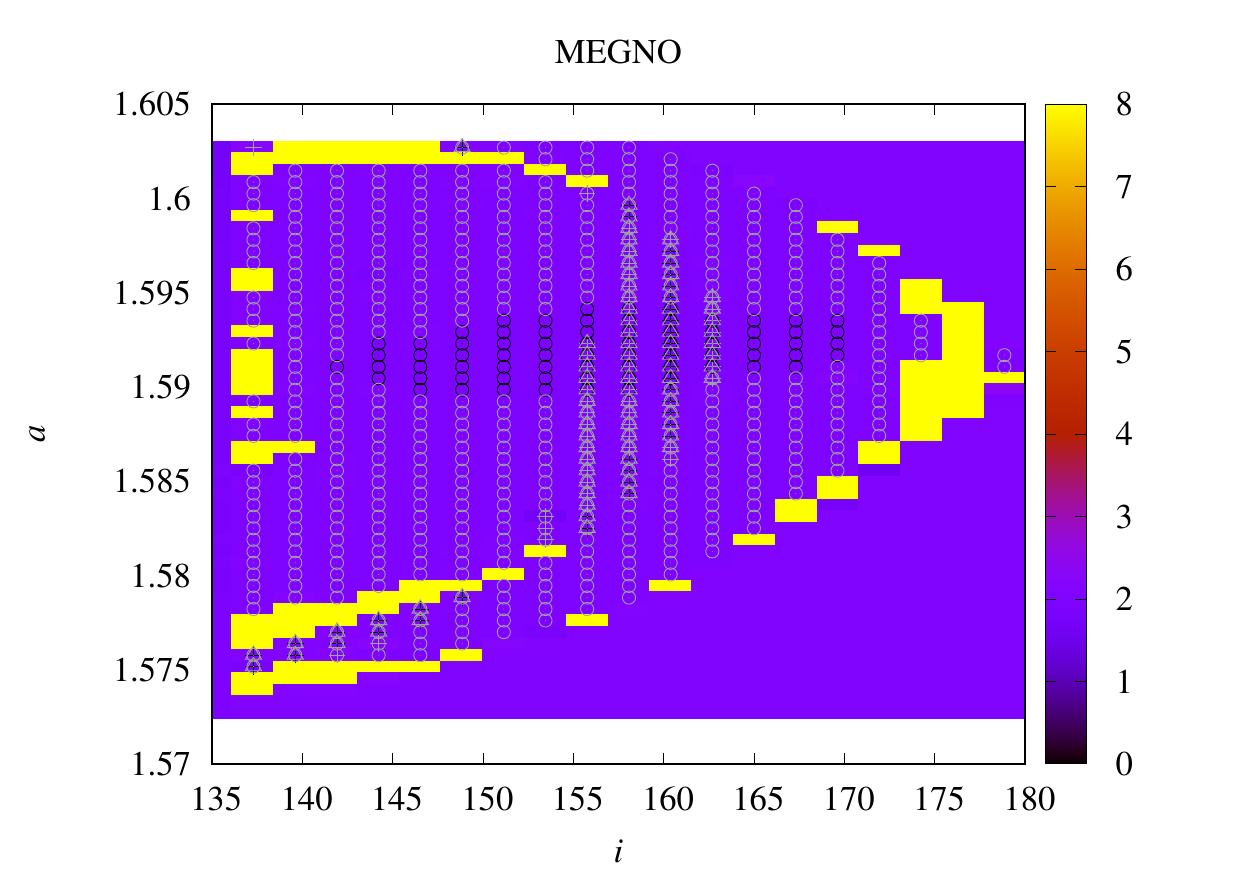}
\includegraphics[width=6.0cm]{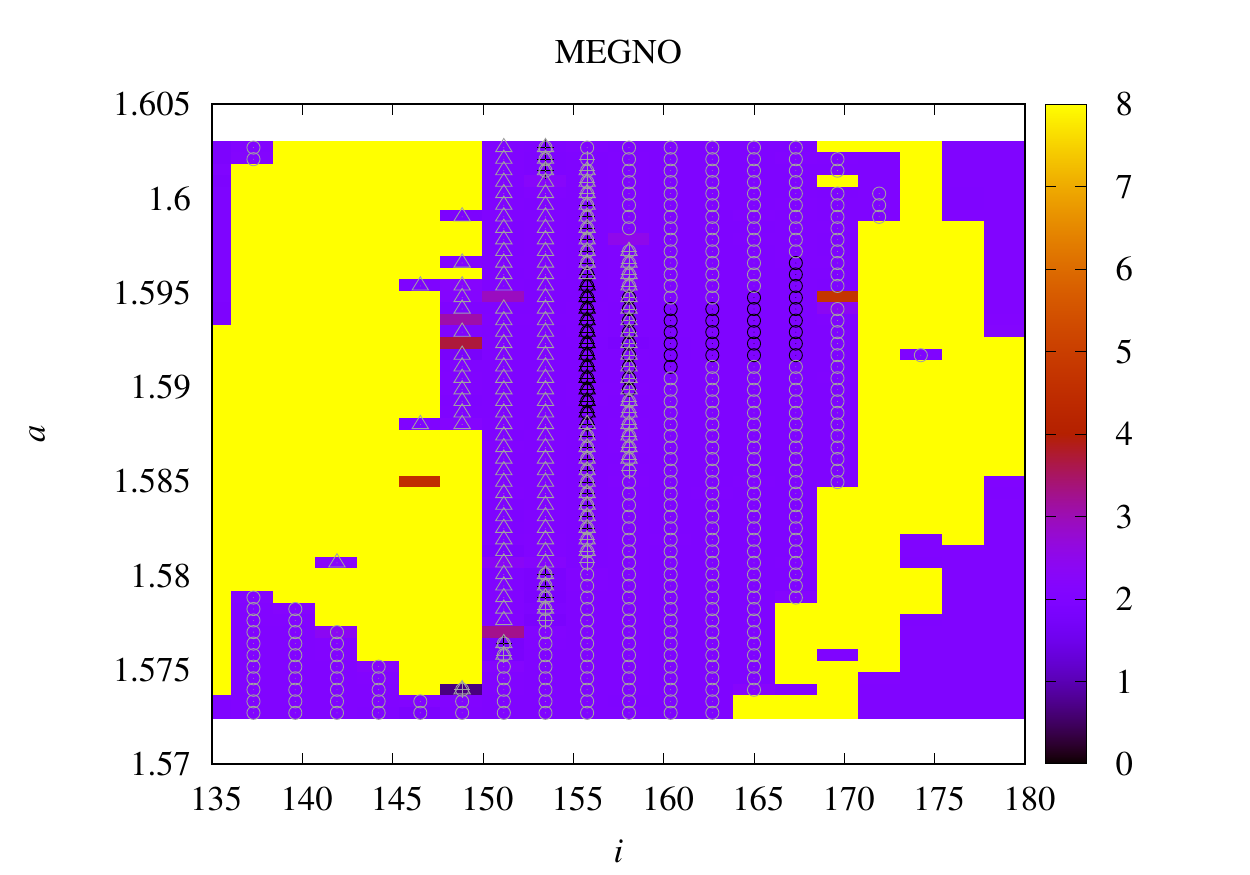}
\includegraphics[width=6.0cm]{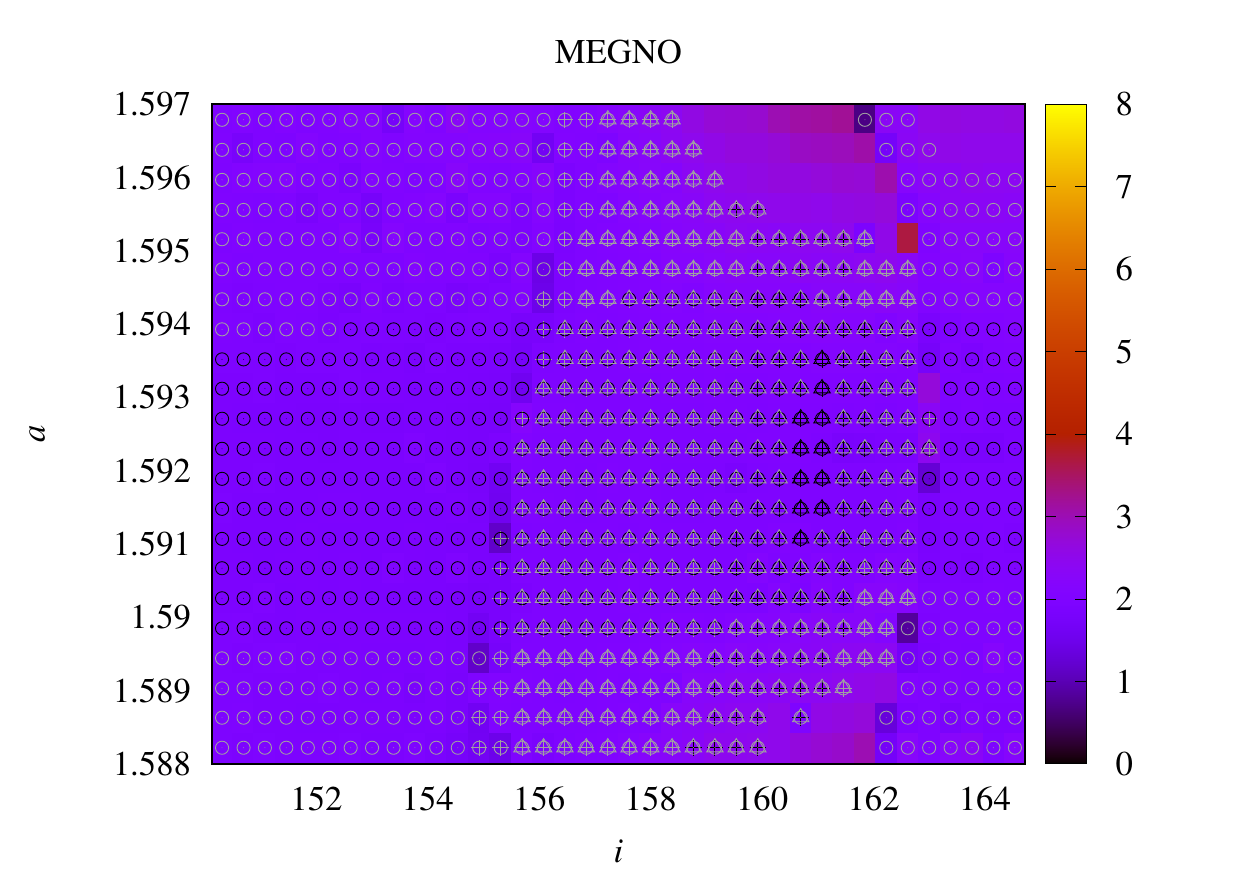}
\includegraphics[width=6.0cm]{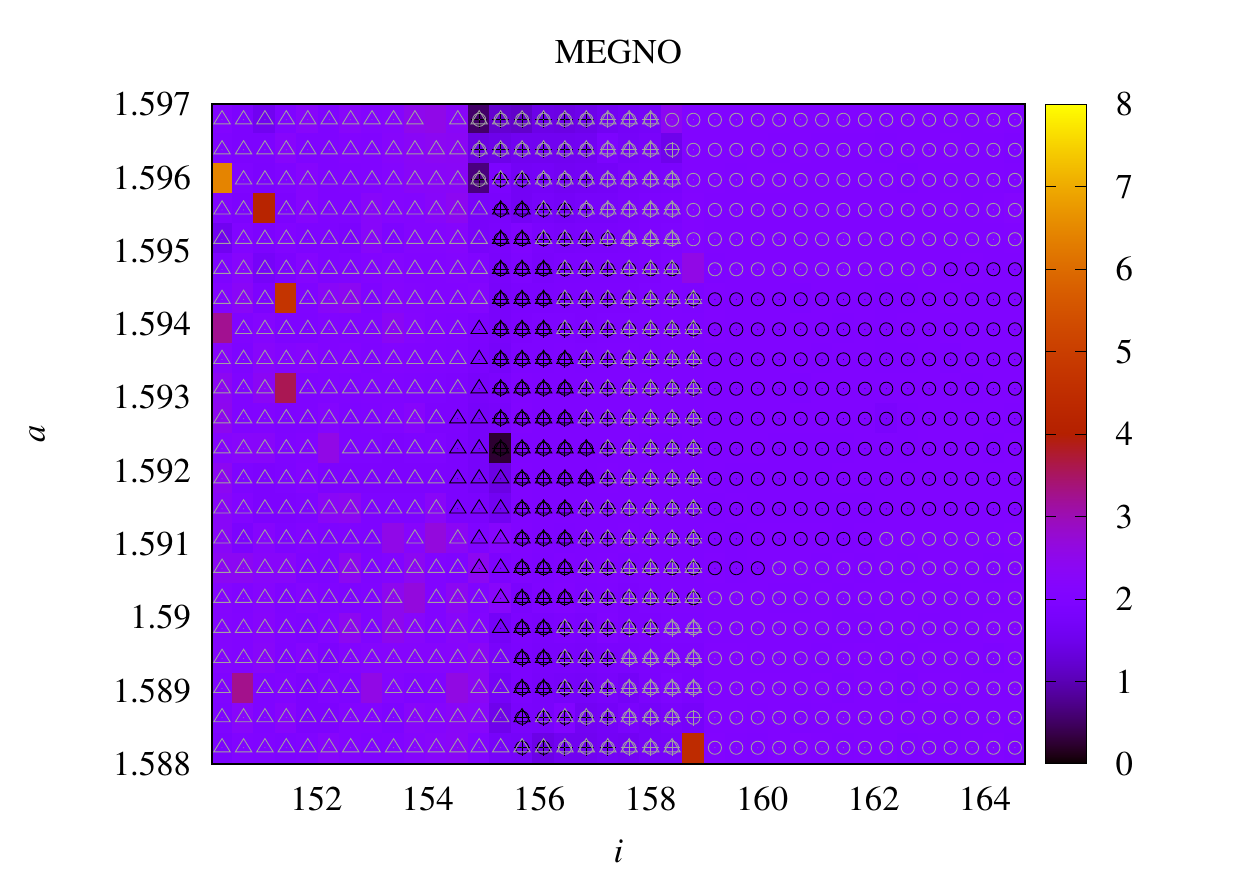}
\caption{MEGNO maps (color scale) for 1/2 retrograde resonance with initial conditions $\omega=90^\circ$, $\phi=180^\circ$ and $\Omega=0$ (thus $\phi^*=0$), $e=0.1$ (left) and $e=0.3$ (right). Symbols indicate libration of the angles $\phi$ around  $180^\circ$ (circles), $\phi^*$  around 0 (triangles) and $\omega$ around $90^\circ$ ($+$ sign). Overlap of circles, triangles and $+$ signs correspond to orbits with  all angles librating (inside the separatrix).   Black (gray) symbols indicate libration amplitude less (larger) than  $50^\circ$.  The lower panels are  zooms of the upper panels  around the fixed points which correspond to the stable branch of the type (A) SPO family.}
\label{9}
\end{figure*}

Additional information on the topology of the 1/2 retrograde resonance may be obtained  by integrating orbits in  $(a,i)$ grids  at set initial values of $e$, $\omega$ and $\phi$. 	In Fig.~\ref{9} we show  examples  of grid integrations over $5\times 10^6$ binary periods with 
$\omega=90^\circ$, $\phi=180^\circ$ (thus $\phi^*=0$), $e=0.1$ (left) and $e=0.3$ (right).  The lower panels show zooms of the upper panels around the fixed points which correspond to the stable branch of the type (A) SPO.  The regions of  $\phi$ libration around $180^\circ$ (marked with circles) and $\phi^*$ libration around 0 (marked with triangles) connect.  The type (A) SPO (marked by the overlap of circles, triangles and $+$ signs, better observed in the lower panels)  is surrounded by  quasiperiodic orbits such that  $\phi$ librates around $180^\circ$. Due to the proximity between the  two SPO families, the set of quasiperiodic orbits  with $\phi$ librating around $180^\circ$ such that $\omega$ circulates is dominant over the set where $\omega$ librates around the Kozai center $90^\circ$. We confirmed numerically that,  when Stoke's drag is included, initial conditions corresponding to these quasiperiodic islands may be captured directly  in the $\phi^*=0$ resonant mode due to the increase in eccentricity which occurs during the $\phi=180^\circ$ mixed-resonance stage.  Note that the stable center corresponding to the type (A) SPO family is  better defined on the right panel ($e=0.3$)  than on the left panel ($e=0.1$) when the stable and unstable families nearly overlap.

\section{Conclusion}

We continued our work on resonant capture in 3D \citep{NamouniMorais2015,NamouniMorais2017,NamouniMorais2018JCAM} and the relation with the PO families \citep{MoraisNamouni2019}.

We obtained the families of POs associated with the 1/2 retrograde resonance at Jupiter to Sun mass ratio. The planar resonant POs correspond to $\phi^*=180^\circ$ which bifurcates from the circular family, and $\phi^*=0$ which  we call the high eccentricity resonant family since it exists only above the pericentric collision line at $e>0.4$.  A family of  type (A) SPOs bifurcates from   the  {\it vcos}  on the circular family and on the resonant $\phi^*=0$ family.  The stable branch of the  type (A) SPO family corresponds to a  stable fixed point of the averaged (resonant) Hamiltonian with $\phi=180^\circ$ and $\omega=90^\circ,270^\circ$ (thus $\phi^*=0$).

 There is a type (B)  SPO family also bifurcating from the {\it vco} on the circular family  which is unstable but  initially nearly critical and corresponds to the libration center $\phi=180^\circ$.  When inclination is close to $180^\circ$ the two SPO families nearly overlap. Due to this proximity, when a small adiabatic dissipation akin to Stoke's drag is included, the orbit follows the  nearly critical type (B) family with $\phi=180^\circ$.  Slow chaotic diffusion on the  type (B) SPO   causes Kozai oscillations around the centers $\omega=90^\circ,270^\circ$ which correspond to the stable branch of the type (A) SPO.   As the semi-major axis decreases due to dissipation these Kozai centers are displaced to larger eccentricity and there is   capture into a quasiperiodic orbit associated with the resonant mode $\phi^*=0$.  This capture is possible once a critical value of the Jacobi constant (associated with a saddle-node bifurcation on the type (A) SPO family) is reached. 

In the case of the 1/2 resonance (studied here) and the 1/1 resonance \citep{MoraisNamouni2019},  3D SPOs bifurcating from the {\it vcos} on the circular family and on the high eccentricty  resonant family connect. In 3D, at small eccentricity, there is libration of the prograde resonant angle $\phi$ around $180^\circ$. When slow dissipation is included these orbits evolve towards  Kozai separatrices.   Kozai libration is extended in the case of the 1/2 resonance as it is associated with bf an unstable but nearly critical  family and not just a single PO as for the 1/1 resonance.
In both cases,  there is chaotic drift on Kozai separatrices and the displacement of the Kozai centers at $\omega=90^\circ,270^\circ$ to larger eccentricity. Capture in a 3D quasiperiodic orbit around the high  eccentricity retrograde resonant mode occurs at a critical point, connecting the stable  and unstable branches bifurcating from the {\it vcos} on the circular and  resonant high eccentricity  families, respectively.

Stable PO families are the likely end states for dynamical systems. However, computation of stable  POs does not provide  information about the extent of quasiperiodic orbits around them. Here,  we showed that in the presence of slow dissipation, unstable but nearly critical POs  mediate transitions between distinct stable families.  
Combining  computation of PO families  with stability maps based on a chaos indicator is useful in order to gain insight into the extent of the quasiperiodic orbits around a PO family and the role played by unstable but nearly critical families which exhibit slow and sticky  chaotic diffusion \citep{zaslavsky07}.

\section*{Acknowledgments}
Bibliography access was provided by CAPES-Brazil.
M.H.M. Morais research had financial support from  S\~ao Paulo Research Foundation (FAPESP/2018/08620-1) and CNPQ-Brazil (PQ2/304037/2018-4) .

\bibliographystyle{mn2e}

\bibliography{biblio}

\end{document}